\documentclass[11pt,twoside,a4paper,cmspaper,final,collab]{cms-tdr}
\def\svnVersion{285b889-D}\def\svnDate{2026/08/04}\def\cmsCernNoTag{CERN-EP-2026-239}\def\cmsCernDate{\today}\def\cmsMessage{Submitted to Reports on Progress in Physics}
\begin{document}\cmsNoteHeader{TOP-25-002}

 \ifthenelse{\boolean{cms@external}}{\providecommand{\cmsLeft}{upper\xspace}}{\providecommand{\cmsLeft}{left\xspace}}
\ifthenelse{\boolean{cms@external}}{\providecommand{\cmsRight}{lower\xspace}}{\providecommand{\cmsRight}{right\xspace}}

\newcommand{\etatxsec}{\ensuremath{5.1\pm0.9\unit{pb}}\xspace}
\newcommand{\etatsigexp}{\ensuremath{7.3}\xspace}
\newcommand{\etatsigobs}{\ensuremath{6.1}\xspace}
\newcommand{\mufitobs}{\ensuremath{0.80\,^{+0.14}_{-0.13}}\xspace}

\newcommand{\ttnrqcdxsec}{\ensuremath{2.8\pm0.5\unit{pb}}\xspace}
\newcommand{\ttnrqcdsigexp}{\ensuremath{13}\xspace}
\newcommand{\ttnrqcdsigobs}{\ensuremath{5.9}\xspace}
\newcommand{\ttnrqcdmufitobs}{\ensuremath{0.44\,^{+0.08}_{-0.07}}\xspace}

\newcommand{\Jbl}{\ensuremath{\PQb_{\Pell}}\xspace}
\newcommand{\MGHPYT}{\MGvATNLO{}+\PYTHIA}
\newcommand{\MINNLO}{\textsc{minnlo}\xspace}
\newcommand{\PWGHRW}{\POWHEG{}+\HERWIG}
\newcommand{\Mtop}{\ensuremath{m_{\PQt}}\xspace}
\newcommand{\PWGPYT}{\POWHEG{}+\PYTHIA}
\newcommand{\SNN}{\ensuremath{S_{\text{NN}}}\xspace}
\newcommand{\Shigh}{\ensuremath{S_{\text{high}}}\xspace}
\newcommand{\Slow}{\ensuremath{S_{\text{low}}}\xspace}
\newcommand{\boost}{\ensuremath{\beta_{\PQt}^{\ast}}\xspace}
\newcommand{\chel}{\ensuremath{c_{\text{hel}}}\xspace}
\newcommand{\chan}{\ensuremath{c_{\text{han}}}\xspace}
\newcommand{\etat}{{\HepParticle{\PGh}{\PQt}{}}\xspace}
\newcommand{\ttnrqcd}{\ensuremath{\ttbar_{\text{GFRW}}}\xspace}
\newcommand{\lpj}{\ensuremath{\Pe/\Pgm\text{+jets}}\xspace}
\newcommand{\muf}{\ensuremath{\mu_{\mathrm{f}}}\xspace}
\newcommand{\mur}{\ensuremath{\mu_{\mathrm{r}}}\xspace}
\newcommand{\pp}{\ensuremath{\Pp\Pp}\xspace}
\newcommand{\tbarpt}{\ensuremath{\pt(\PAQt)}\xspace}
\newcommand{\topy}{\ensuremath{\abs{y(\PQt)}}\xspace}
\newcommand{\tpt}{\ensuremath{\pt(\PQt)}\xspace}
\newcommand{\tqh}{\ensuremath{\PQt_{\mathrm{h}}}\xspace}
\newcommand{\tql}{\ensuremath{\PQt_{\Pell}}\xspace}
\newcommand{\ttm}{\ensuremath{m(\ttbar)}\xspace}
\newcommand{\yt}{\ensuremath{Y_{\PQt}}\xspace}
\newcommand{\hdamp}{\ensuremath{h_{\text{damp}}}\xspace}
\newcommand{\Wadd}{\ensuremath{\PW_{\text{add}}}\xspace}
\newcommand{\DYadd}{\ensuremath{\text{DY}_{\text{add}}}\xspace}
\newcommand{\MCdiff}{\ensuremath{\text{MC}_{\text{diff}}}\xspace}

\cmsNoteHeader{TOP-25-002}
\title{Observation of an excess at the top quark pair production threshold in the single-lepton channel}

\author[cern]{CMS Collaboration}

\date{\today}

\abstract{
   A search is presented for top quark-antiquark (\ttbar) quasi-bound ``toponium'' states, near the kinematic \ttbar production threshold in final states with a single electron or muon and jets. The study uses proton-proton collision data at $\sqrt{s}=13\TeV$, collected by the CMS experiment at the CERN LHC, corresponding to an integrated luminosity of 138\fbinv. The analysis examines the relative velocity between the top quark and antiquark, along with two angular observables sensitive to the parity and spin of the \ttbar system. A significant excess of events is observed relative to the standard model prediction for \ttbar production calculated at next-to-next-to-leading order in perturbative quantum chromodynamics (QCD). The excess corresponds to an observed total cross section of \etatxsec and is consistent with a simplified model of a color singlet pseudoscalar toponium motivated by nonrelativistic QCD, with a resulting observed significance of \etatsigobs standard deviations. The result provides an independent confirmation of the excess reported in the dilepton channel.
}

\hypersetup{%
pdfauthor={CMS Collaboration},%
pdftitle={Observation of an excess at the top quark pair production threshold in the single-lepton channel},%
pdfsubject={CMS},%
pdfkeywords={CMS, top, toponium, quasi-bound states, production threshold, single-lepton channel}}

\maketitle

\section{Introduction}

As the heaviest known fundamental particle, the top quark decays with a mean lifetime of \mbox{$5\times10^{-25}\unit{s}$}, which is shorter than the quantum chromodynamics (QCD) hadronization time\-scale $1/\Lambda_{\text{QCD}}\approx10^{-24}\unit{s}$~\cite{ParticleDataGroup:2024}. This short lifetime implies that, unlike lighter quarks, the top quark typically decays before hadronization can occur.

Nevertheless, in proton-proton (\pp) collisions at the CERN LHC, a small contribution from quasi-bound states in the production of top quark-antiquark pairs (\ttbar), here referred to as toponium, is still predicted according to calculations in nonrelativistic QCD (NRQCD)~\cite{BIGI1986157, Fadin:1990wx, Kiyo:2008bv, Hagiwara_2008, Sumino:2010bv, Ju:2020otc, Fuks:2021xje, Garzelli:2024uhe, fuks2024}. Toponium is expected to predominantly exist in a pseudoscalar color singlet state, \etat, which would produce an excess in the invariant mass distribution of the \ttbar system (\ttm) just below the continuum production threshold of twice the top quark mass (\Mtop). Both the ATLAS and CMS Collaborations have reported an excess of events in this region relative to the standard model (SM) prediction for \ttbar production, calculated at next-to-next-to-leading order (NNLO) in perturbative QCD~\cite{TOP-24-007, ATLAS-CONF-2025-008}.

The \etat decays predominantly through the weak decay of its constituent top quarks, each producing a \PW boson and a bottom quark. The \PW boson decays either hadronically into a pair of quarks, which subsequently hadronize into jets, or leptonically into a charged lepton and a neutrino.

So far, the excess has only been observed in the dilepton channel, where both \PW bosons decay leptonically. This analysis focuses on the \lpj channel, where one \PW boson decays leptonically and the other hadronically, excluding events with electrons or muons from the decay of tau leptons originating from top quarks. The presented measurement utilizes data collected by the CMS experiment between 2016 and 2018, with a corresponding integrated luminosity of 138\fbinv.

The \lpj channel offers two main advantages over the dilepton channel: a greater number of events due to a branching fraction of about six times that of the dilepton channel, when not accounting for tau leptons~\cite{ParticleDataGroup:2024}, and simpler kinematic reconstruction because of the presence of only one neutrino instead of two. Despite these advantages, establishing the \etat signal in this channel remains challenging. The reconstruction of \ttm in final states with four or more jets is accompanied by significant systematic uncertainties, particularly those related to jet energy corrections. Although the signal excess is clearly pronounced in theoretical calculations~\cite{Kiyo:2008bv}, it becomes significantly diluted once parton showering and hadronization effects are included, and is further smeared by detector resolution effects. Additionally, the simulated \ttm distribution near the threshold is strongly influenced by the assumed \Mtop value, which constitutes a major source of uncertainty in the \etat search.

To mitigate these effects, we use the relative velocity of the top quarks as an alternative to \ttm, as it is more sensitive to the \etat signal and more robust against the aforementioned uncertainties. In addition, following the dilepton analyses~\cite{TOP-24-007, ATLAS-CONF-2025-008}, we exploit angular distributions of the \ttbar decay products that provide sensitivity to the spin correlations and parity of the \ttbar system. In the dilepton channel, these observables can be calculated using the two charged leptons. However, in the \lpj channel, the lower efficiency in the identification of the corresponding down-type quark from the hadronic \PW boson decay reduces the sensitivity of the angular observables.

The remainder of this paper is organized as follows. In Section~\ref{sec:strategy}, we discuss the choice of variables and the analysis strategy. The CMS detector is described in Section~\ref{sec:cmsdetector}. The details of signal and background modeling are presented in Section~\ref{sec:samples}, followed by the description of the reconstruction of physics objects in Section~\ref{sec:objects} and event reconstruction in Section~\ref{sec:reconstruction}. The estimation of the background contribution from sources other than top quark pair production is discussed in Section~\ref{sec:background}. In Section~\ref{sec:signalfit}, we detail the statistical model used to extract the signal. Sources and treatment of systematic uncertainties are described in Section~\ref{sec:uncertainties}. Results are presented in Section~\ref{sec:results}, followed by the summary in Section~\ref{sec:summary}. Tabulated results are provided in the HEPData record for this analysis~\cite{hepdata}.

\section{Analysis strategy}
\label{sec:strategy}

In this Section, we introduce the sensitive observables used in this measurement. We denote the top quark with the leptonic (hadronic) \PW boson decay by \tql (\tqh). The quantity \boost is defined as the magnitude of the relative velocity between the two top quarks:
\begin{equation}
    \boost = \abs{\mathbf{p}}/E,
\end{equation}
where $\mathbf{p}$ and $E$ are the momentum and energy of \tql measured in the rest frame of \tqh, respectively. This variable serves as an alternative to \ttm. The relation between \ttm and \boost is given by
\begin{equation}\label{eq:boost}
    \frac{m^2(\ttbar)-m^2(\tql)-m^2(\tqh)}{2m(\tql)m(\tqh)} = \frac{1}{\sqrt{1-{\boost}^2}}.
\end{equation}
The variables $\abs{\mathbf{p}}$ and $E$ have similar energy dependencies, meaning systematic uncertainties such as jet energy scale variations affect them in a comparable way. Since \boost is a ratio of these two variables, it is significantly more robust against those effects. Furthermore, examining the left-hand side of Eq.~\eqref{eq:boost}, one can see that instead of measuring \ttm directly, \boost measures the deviation of $m^2(\ttbar)$ from $m^2(\tql)+m^2(\tqh)$, making it less sensitive to the exact value of \Mtop, and allowing for a partial cancellation of resolution effects shared by the reconstructed \ttm and top quark masses.

To probe the spin correlations and parity of the \ttbar system, we follow the methodology of Ref.~\cite{CMS-PAS-HIG-22-013} and introduce two angular observables
\begin{equation}
    \chel = \widehat{\Pell}\cdot\widehat{\PQd}
    \quad\text{and}\quad
    \chan = \widehat{\Pell}{}^{\prime}\!\cdot\widehat{\PQd},
\end{equation}
where $\widehat{\Pell}$ and $\widehat{\PQd}$ denote the unit momentum vectors of the charged lepton and the down-type (anti)quark from the \PW boson decay, respectively. Each vector is evaluated in the rest frame of its parent top (anti)quark, using the helicity basis defined in the \ttbar rest frame following Ref.~\cite{Bernreuther:2015yna}. The $\widehat{\Pell}{}^{\prime}$ vector is obtained from $\widehat{\Pell}$ by reversing its component along the top quark direction defined in this basis. The charged leptons and the down-type quarks are selected because they carry the highest amount of information about the top (anti)quark spins~\cite{Brandenburg:2002xr}. The strategy for distinguishing up- and down-type quarks is described in Section~\ref{sec:reconstruction}.

Finally, the signal strength is extracted from fitting the predicted contributions from \etat production, continuum \ttbar, and non-\ttbar background processes to the observed data distribution in these variables.

\section{The CMS detector}
\label{sec:cmsdetector}

The CMS detector~\cite{CMS:2008xjf,CMS:2023gfb} comprises several highly specialized components. At its core is a superconducting solenoid of 6\unit{m} internal diameter, generating a 3.8\unit{T} magnetic field. Inside the solenoid are key subsystems: a silicon pixel and strip tracker, a lead tungstate scintillating-crystals electromagnetic calorimeter (ECAL), and a brass and scintillator sampling hadron calorimeter (HCAL). These are arranged into a central barrel region with two endcaps, while additional forward calorimeters extend the pseudorapidity ($\eta$) coverage. Outside the solenoid is the steel flux-return yoke, which has layers of gas-ionization chambers embedded within to reconstruct muon trajectories.

The CMS experiment manages the data from collisions with a sophisticated two-level trigger system. The first level uses custom hardware processors to quickly analyze signals from the calorimeters and muon detectors, selecting events at roughly 100\unit{kHz} within a fixed latency of about 4\mus~\cite{CMS:2020cmk}. Selected events are then processed by a high-level trigger---a farm of processors running a full event reconstruction software optimized for fast processing---that further reduces the event rate to a few kHz before data storage~\cite{CMS:2016ngn,CMS:2024aqx}. For this analysis, event selection is performed using single-electron and -muon triggers for isolated leptons. The minimum transverse momentum (\pt) thresholds for these triggers vary by year, with values set at 24 or 27\GeV for muons, and 27 or 32\GeV for electrons.

{\tolerance=800
To accurately reconstruct the physics objects within each recorded event, CMS employs a particle-flow (PF) algorithm~\cite{PF} that integrates information from all detector components, aiming to reconstruct and identify each particle in an event. In this approach, the energy of photons is directly measured in the ECAL. An electron's energy is reconstructed by considering the momentum measured at the primary vertex, the energy recorded in the corresponding ECAL cluster, and the summed energy of all bremsstrahlung photons that are spatially compatible with the electron's track. The primary vertex is defined as the vertex corresponding to the hardest scattering in the event, determined solely from tracking information (see Section~9.4.1 of Ref.~\cite{CMS-TDR-15-02}). A muon's momentum is determined from the curvature of its tracks, while the energy of a charged hadron is derived from a combination of the momentum measured in the tracker and the matching energy deposits in ECAL and HCAL, adjusted for the calorimeters' response functions to hadronic showers. Finally, the energy of a neutral hadron is identified through the corresponding corrected energies from both ECAL and HCAL. For a comprehensive overview of the CMS detector, including details on its coordinate system and key kinematic variables, please refer to Refs.~\cite{CMS:2008xjf,CMS:2023gfb}.
\par}

\section{Signal and background modeling}
\label{sec:samples}

Signal and background events are modeled using a variety of Monte-Carlo (MC) simulation programs. For the \etat simulation, we adopt a simplified model based on Refs.~\cite{Fuks:2021xje, Maltoni:2024tul}, using a configuration identical to that of the CMS dilepton analysis~\cite{TOP-24-007} to ensure direct comparability. This involves simulating a color singlet pseudoscalar particle with a mass of 343\GeV and a width of 2.8\GeV, corresponding to approximately twice the top quark width~\cite{ParticleDataGroup:2024}. The total cross section for \etat production is assumed to be 6.43\unit{pb} as indicated in Ref.~\cite{Fuks:2021xje}. The matrix element (ME) event generator \MGvATNLO2.6.1~\cite{Alwall:2014hca} is used to simulate \etat production with leading order (LO) QCD accuracy. The particle is produced via gluon-gluon fusion and decays as $\etat\to\PW\PQb\PW\PAQb$. This is then combined with the parton shower (PS) and hadronization simulation from \PYTHIA8.240~\cite{Sjostrand:2014zea}. Unless otherwise specified, all \PYTHIA simulations in this analysis are performed using the CP5 underlying event tune~\cite{CP5}.

An alternative model, proposed in Ref.~\cite{fuks2024}, evaluates NRQCD effects in S-wave color singlet \ttbar production using the Green's function approach in the Coulomb gauge. Near the production threshold, \ttbar events obtained at LO QCD with \MGvATNLO are reweighted to match the NRQCD calculation as a function of \ttm and the top quark momentum in the \ttbar rest frame. The reweighting is applied exclusively to events with $\ttm<350\GeV$ and a top quark momentum of less than 50\GeV. The predictions are interfaced with \PYTHIA. The current calculation of the Green's function is obtained with $\Mtop=173\GeV$, which slightly differs from the $\Mtop=172.5\GeV$ used for all other top quark production samples in this analysis. This Green's function reweighted \ttbar, hereafter denoted as \ttnrqcd, serves as an alternative signal model for a direct comparison to Ref.~\cite{ATLAS-CONF-2025-008}. The double-counting correction described in Ref.~\cite{fuks2024}, which accounts for the overlap between the NRQCD threshold contribution and the fixed-order continuum production, is omitted to maintain alignment with Ref.~\cite{ATLAS-CONF-2025-008}. The assumed total cross section for the \ttnrqcd contribution is taken to be the same as that of \etat.

The main background in this analysis is continuum \ttbar events, where the pairs do not form a resonant state. The simulation of these events is carried out using the \POWHEG \textsc{v2} ME event generator~\cite{Nason:2004rx, Frixione:2007vw, Frixione:2007nw, Alioli:2010xd} with next-to-LO (NLO) QCD accuracy, interfaced with \PYTHIA. The sample is then reweighted using a neural network (NN)-based method~\cite{mlreweighting, CMS-DP-2023-031} that uses the full kinematic information in each event to account for NNLO QCD effects, which are obtained with \POWHEG \MINNLO~\cite{Mazzitelli_2021} and interfaced with \PYTHIA. Contributions of electroweak (EW) corrections are estimated using the \textsc{hathor} \textsc{2.1-b3} package~\cite{hathorart} and are applied as $k$-factors that are functions of \ttm and the top quark scattering angle in the \ttbar rest frame:
\begin{equation}\label{eq:Kfac}
    k^{\text{NLO}}_{\text{EW}}(\yt) = \frac{\sigma_{\text{NLO EW}}(\yt)}{\sigma_{\text{LO QCD}}},
\end{equation}
where $\sigma_{\text{NLO EW}}$ and $\sigma_{\text{LO QCD}}$ represent the continuum \ttbar contributions at NLO EW and LO QCD precision, respectively, and \yt is the top quark Yukawa coupling, set to unity as the default. All \ttbar samples, including the \etat and \ttnrqcd simulations, are generated for all decay channels of the top quarks.

\begin{figure*}[!ht]
  \centering
  \includegraphics[width=0.495\textwidth]{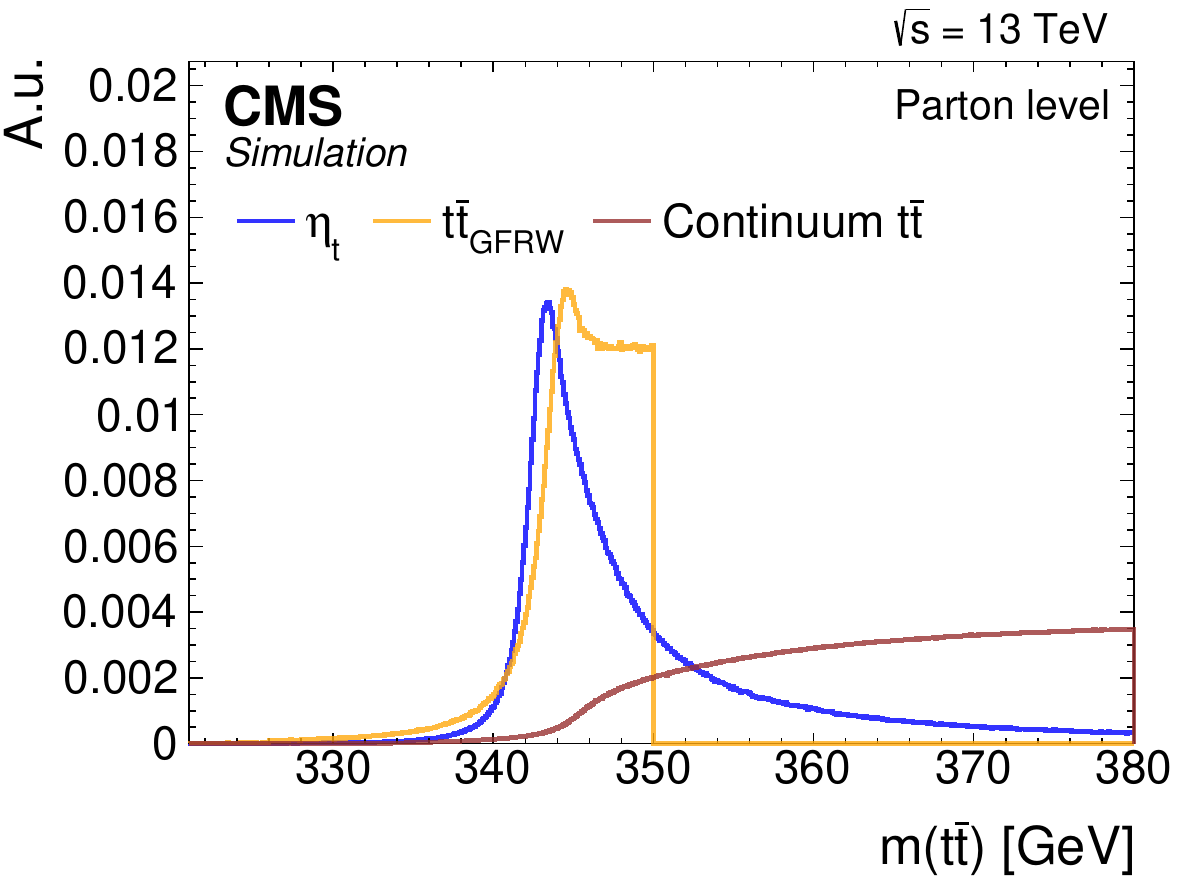}%
  \hfill%
  \includegraphics[width=0.495\textwidth]{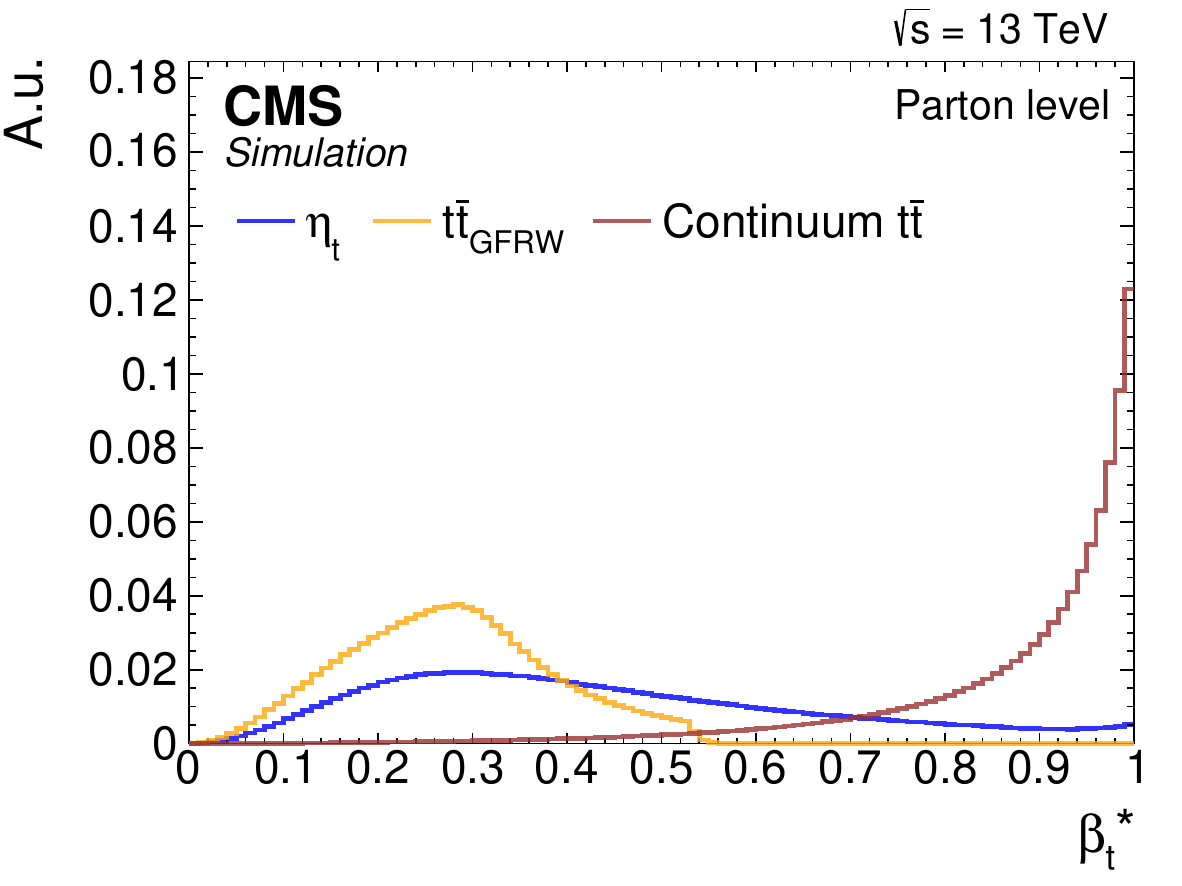} \\
  \includegraphics[width=0.495\textwidth]{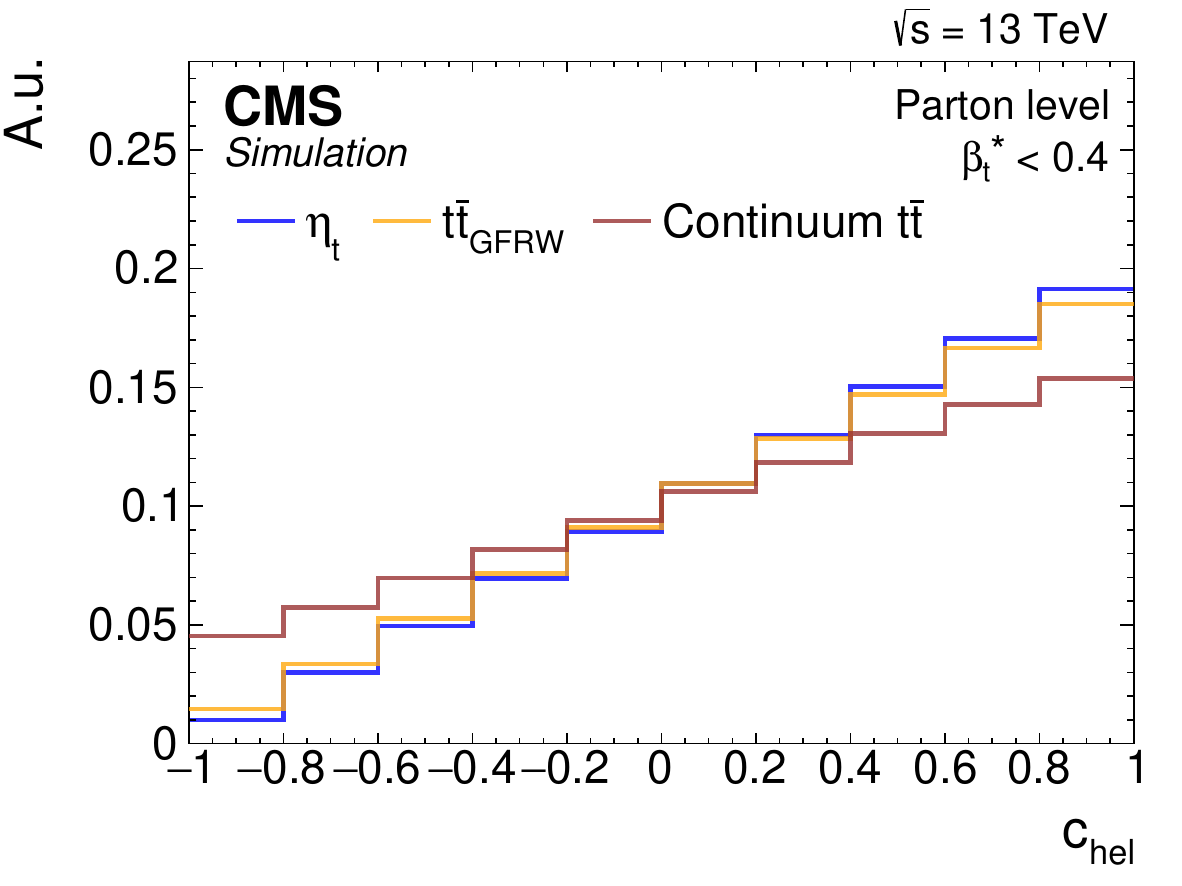}%
  \hfill%
  \includegraphics[width=0.495\textwidth]{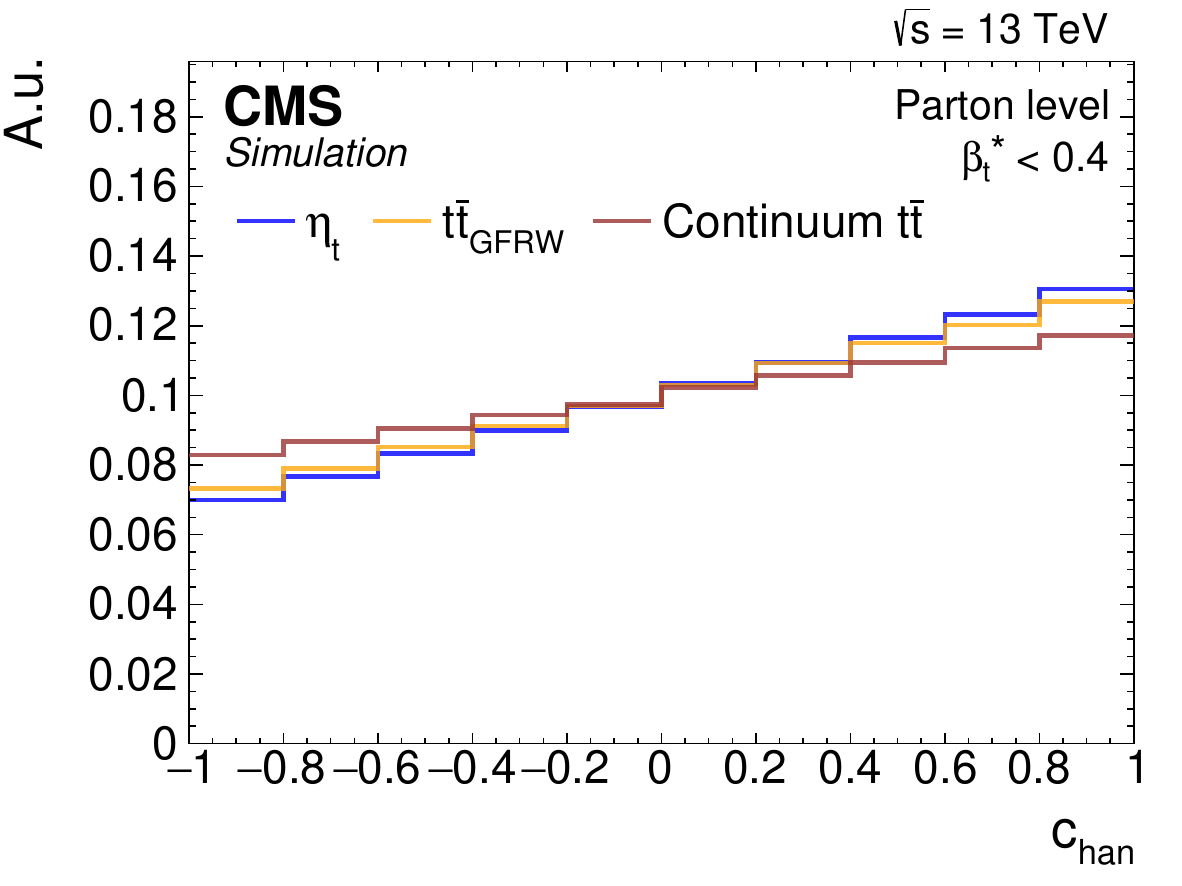}
  \caption{Parton-level distributions of \ttm (upper left), \boost (upper right), \chel (lower left), and \chan (lower right) for the \etat model (blue), the \ttnrqcd model (orange), and the continuum \ttbar background (red). The \chel and \chan distributions are restricted to events with $\boost<0.4$ at the parton level to focus on the threshold region. The distributions for the continuum \ttbar background include the NNLO QCD reweighting and the EW corrections. All distributions are normalized to unit area.}
  \label{fig:signal_compare}
\end{figure*}

A comparison between the simplified \etat model, the \ttnrqcd model, and the continuum \ttbar background is presented in Fig.~\ref{fig:signal_compare}, displaying the normalized parton-level distributions for \ttm, \boost, \chel, and \chan. The distributions for continuum \ttbar events include the NNLO QCD reweighting and the EW corrections. The \ttnrqcd model differs significantly from the \etat model in the \ttm distribution, primarily due to the kinematic cutoff at 350\GeV, the omission of the double-counting correction resulting in a plateau near the cutoff, and the difference in the \Mtop assumption shifting the peak. In the \boost distribution, the continuum production peaks at one because most events are well above the threshold, resulting in a high relative velocity between the top quarks. Meanwhile, the \ttnrqcd model is more populated in the lower bins than the \etat model due to the 350\GeV cutoff. Both signal models exhibit similar behavior in the angular observables, which have more positive slopes than the continuum \ttbar due to the anticorrelated spins of the top quarks~\cite{CMS-PAS-HIG-22-013}.

The pre-fit normalization of the continuum \ttbar samples is based on the inclusive production cross section of $832\,^{+40}_{-46}\unit{pb}$, determined at NNLO QCD accuracy with soft-gluon resummation at the next-to-next-to-leading logarithmic level~\cite{Czakon:2011xx}. For the \etat, \ttnrqcd, and NLO QCD \ttbar simulations, the renormalization (\mur) and factorization (\muf) scales are both set to the average transverse mass of the top quark and antiquark, expressed as $\mT = \big(\sqrt{\smash[b]{\Mtop^2+\tpt^2}} + \sqrt{\smash[b]{\Mtop^2+\tbarpt^2}}\big)/2$. Here, the \pt of the top quark is evaluated in the \ttbar rest frame and $\Mtop=172.5\GeV$~\cite{CMS:2015lbj} is used throughout this paper unless otherwise specified. For the \ttbar events simulated with \POWHEG \MINNLO, \mur and \muf are set to $\mT/2$ as recommended in Ref.~\cite{NNLOscale}. The uncertainties in the \ttbar modeling, including those arising from the choice of \mur, \muf, and \Mtop, are evaluated as described in Section~\ref{sec:uncertainties}.

{\tolerance=2400
Non-\ttbar background contributions include single top quark production, Drell--Yan (DY) processes, \PW boson production in association with jets, and events consisting solely of jets produced via QCD interactions, known as multijet events. Single top quark production via the $t$ channel and top quark production in association with a \PW boson are simulated using \PWGPYT. Meanwhile, \MGHPYT is used to simulate single top quark production via the $s$ channel at NLO QCD. The simulation of DY and \PW boson backgrounds is performed at LO QCD with \MGvATNLO{}+\PYTHIA, incorporating MLM PS matching~\cite{Mangano:2006rw, MLM} of up to four partons. For the multijet processes, \PYTHIA is used at LO. The \PW boson, DY, and single top quark backgrounds are normalized to their corresponding inclusive cross sections, calculated at NNLO QCD accuracy for \PW boson and DY~\cite{Li:2012wna}, and at NLO QCD for the single top quark production~\cite{Kant:2014oha, Kidonakis:2012rm}. All simulations use the NNLO version of NNPDF 3.1~\cite{NNPDF31} parton distribution functions (PDFs).
\par}

{\tolerance=10000
The detector response is modeled using \GEANTfour~\cite{Agostinelli:2002hh}. Pileup, which consists of additional \pp interactions that occur within the same or neighboring bunch crossings, is simulated by overlaying additional minimum-bias \pp interactions generated with \PYTHIA, assuming a total inelastic \pp cross section of 69.2\unit{mb}~\cite{CMS:2018mlc}. The simulated events are then reweighted to reproduce the distribution of the number of pileup interactions expected from the instantaneous luminosity profile observed in data. The same reconstruction algorithms used for the data are also applied to simulated events.
\par}

\section{Physics object reconstruction}
\label{sec:objects}

This analysis relies on accurately reconstructing and identifying electrons, muons, jets, and the missing transverse momentum \ptvecmiss associated with neutrinos.

The selection criteria for electrons~\cite{CMSelectrons} and muons~\cite{Sirunyan:2018} are as follows: they must pass the ``tight'' working point defined in Refs.~\cite{CMSelectrons, Sirunyan:2018}, have $\pt>30\GeV$, and $\abs{\eta}<2.4$. For the 2018 data set, the electron \pt is required to be greater than 34\GeV because of an increase in the trigger threshold. Additionally, leptons must be isolated and compatible with originating from the primary vertex. The efficiencies for electron and muon reconstruction and selection are determined using the ``tag-and-probe'' technique~\cite{TNP}. These efficiencies depend on \pt and $\eta$, with electrons having an overall efficiency of 50--80\%, and muons of 75--85\%.

All PF candidates are clustered into jets using the anti-\kt jet algorithm provided by the \FASTJET package~\cite{Cacciari:2008gp, Cacciari:2011ma}, with a distance parameter of 0.4. To minimize pileup effects, charged candidates originating from pileup interaction vertices are removed before clustering. Additionally, a multivariate technique is employed to reject pileup jets, using the loose pileup jet identification (PU jet ID) criteria from Ref.~\cite{puId} for jets with $\pt<50\GeV$. After the jets are formed, the total jet energy is corrected to account for energy depositions from pileup, and further corrections dependent on \pt and $\eta$ are applied to adjust for the detector response~\cite{JET}. Additionally, to resolve ambiguities between jet and lepton candidates, a jet is discarded from the analysis if it has an isolated lepton with $\pt>15\GeV$ within a cone of $\DR = \sqrt{\smash[b]{(\Delta\phi)^2+(\Delta\eta)^2}} = 0.4$, where $\phi$ is the azimuthal angle. Only jets with $\pt>25\GeV$ and $\abs{\eta}<2.4$ are ultimately selected for further analysis.

To identify \PQb jets, we employ the \textsc{DeepJet} algorithm~\cite{Bols_2020, BTV-16-002}. This technique uses an NN that calculates a discriminant value for each jet, which helps separate \PQb jets from those of other flavors. Each jet's discriminant score is compared against three preset thresholds (tight, medium, and loose), and the jet is categorized according to the highest threshold that its score exceeds. The tight threshold identifies \PQb jets with an efficiency of about 50--70\%, depending on the jet \pt and $\eta$. Similarly, the medium and loose thresholds achieve efficiencies of approximately 70--82\% and 85--92\%, respectively. In terms of rejecting non-\PQb jets, the tight, medium, and loose thresholds eliminate around 97, 85, and 55\% of \PQc jets, and about 99.5, 98, and 90\% of light-flavor quark and gluon jets, respectively.

In each event, we calculate the \ptvecmiss, whose magnitude is labeled \ptmiss~\cite{CMS:2019ctu}, by summing the \pt vectors of all PF candidates and then taking the negative of that total. Additionally, this vector is adjusted to incorporate the jet energy scale corrections applied to the reconstructed jets.

Data were collected from 2016 to 2018, with each year analyzed using simulation sets and correction factors that reflect the specific data-taking conditions and detector configurations at the time. Notably, significant changes in the detector configuration affecting tracking efficiency in 2016 required two distinct sets of simulations and correction factors. Consequently, the analysis considers four separate data-taking periods.

\section{Event reconstruction}
\label{sec:reconstruction}

To reconstruct the \ttbar system in the \lpj channel, we follow the method outlined in Ref.~\cite{spincorr}, which is detailed in this Section. We first require events to feature exactly one selected electron or muon (with no additional isolated electron or muon having $\pt>15\GeV$) and at least four jets. In the \ttbar simulation, we associate detector-level objects to parton-level quarks or leptons within $\DR<0.2$. If multiple parton-level candidates are found, the one with the highest \pt is selected. When every particle from the top quark decays at the parton level can be uniquely associated with a detector-level counterpart, the event is labeled ``reconstructible''; otherwise, it is considered ``nonreconstructible''. Because the simulated \ttbar sample is inclusive in all decay modes, the selection includes events with a tau lepton in top quark decays and events from the dilepton and all-jets channels, which are categorized as ``non \lpj'' events.

To match the detector-level objects to the corresponding particles produced in the decays of \tql and \tqh, we rely on an NN that receives a vector encoding the kinematic and \PQb tagging information of the reconstructed objects. The input begins with the four-momentum components ($p_x$, $p_y$, $p_z$, $E$) of the electron or muon, followed by the two components of \ptvecmiss ($p^{\text{miss}}_{x}$, $p^{\text{miss}}_{y}$), and finally, information for up to eight jets, including their four-momentum components and the \PQb tagging category introduced in Section~\ref{sec:objects} ($p_x$, $p_y$, $p_z$, $E$, \PQb cat.). The NN architecture consists of seven fully connected layers of 220 nodes each, using the hyperbolic tangent activation function, followed by a single output node with a sigmoid activation function to constrain the output to the $[0,1]$ range. In total, the network comprises roughly 300\,000 parameters.

The NN is trained to recognize when the first four jet positions in the input vector contain the jets from the \ttbar decay in the following order: the \PQb jet from the \tql decay, the \PQb jet from the \tqh decay, and the down- and up-type quarks from the hadronic \PW boson ($\PW_{\mathrm{h}}$) decay, respectively. To cover the various jet assignment possibilities, every event is presented to the NN with all permutations of four jets selected from the available jets (up to eight jets). Specifically, the first four jet positions in the input vector are filled by these permutations, while remaining jets are added in decreasing order of \pt. This results in a range of 24 possibilities for events with four jets to 1680 possibilities for events with eight jets. The network is trained using supervised labels, assigning a target value of one if the jets are correctly assigned (labeled as ``correct''), and zero otherwise (labeled as ``incorrect''). Training is performed using about 20 million \lpj events from the simulated continuum \ttbar samples, following the previously described selection criteria. The training employs a stochastic method with batches of 128 events and uses the Adam algorithm~\cite{adam} to minimize the logistic loss function. The training sample includes nonreconstructible events with a target value of zero. During training, the logistic loss function is tracked using a validation sample. The losses for both the training and validation samples align, with no signs of overfitting observed. During inference, the permutations are constructed following the same procedure as in training and are sequentially input into the NN for each event. The permutation that produces the highest NN score (\SNN) is used.

Neutrino reconstruction proceeds by imposing the \PW boson mass constraint $(p_\PGn+p_\Pell)^2 = m_{\PW}^2$, using \ptvecmiss as the \pt of the neutrino~\cite{CMS_ttdiff21}. This constraint leads to a quadratic equation for the longitudinal component of the neutrino momentum $p_{z}(\PGn)$. If no real solution exists, the $x$ and $y$ components of \ptvecmiss are scaled with minimal alteration to \ptmiss, selecting the point in the transverse plane closest to the original \ptvecmiss for which a solution exists. In cases where two solutions emerge, the one yielding an invariant mass $m(\tql)^2 = (p_\PGn+p_\Pell+p_{\Jbl})^2$ closer to \Mtop is selected. Even though one or both top quarks from \etat decay are slightly off-shell, this selection does not introduce any noticeable bias on the neutrino momentum. The \ptvecmiss, possibly modified to find the solution, together with the calculated $p_{z}(\PGn)$, form the neutrino momentum.

Additional quality selection criteria are imposed to further suppress backgrounds and increase the fraction of correctly reconstructed events. These include mass window requirements for the reconstructed particles, namely, $\abs{m(\tql)-172.5\GeV}<50\GeV$, $\abs{m(\tqh)-172.5\GeV}<50\GeV$, and $\abs{m(\PW_{\mathrm{h}})-80.4\GeV}<30\GeV$. We also discard events with $\SNN<0.1$ due to their low fraction of correctly reconstructed events and the large contribution of background processes. These criteria define the signal region (SR).

\begin{figure}[!ht]
\centering
\includegraphics[width=0.48\textwidth]{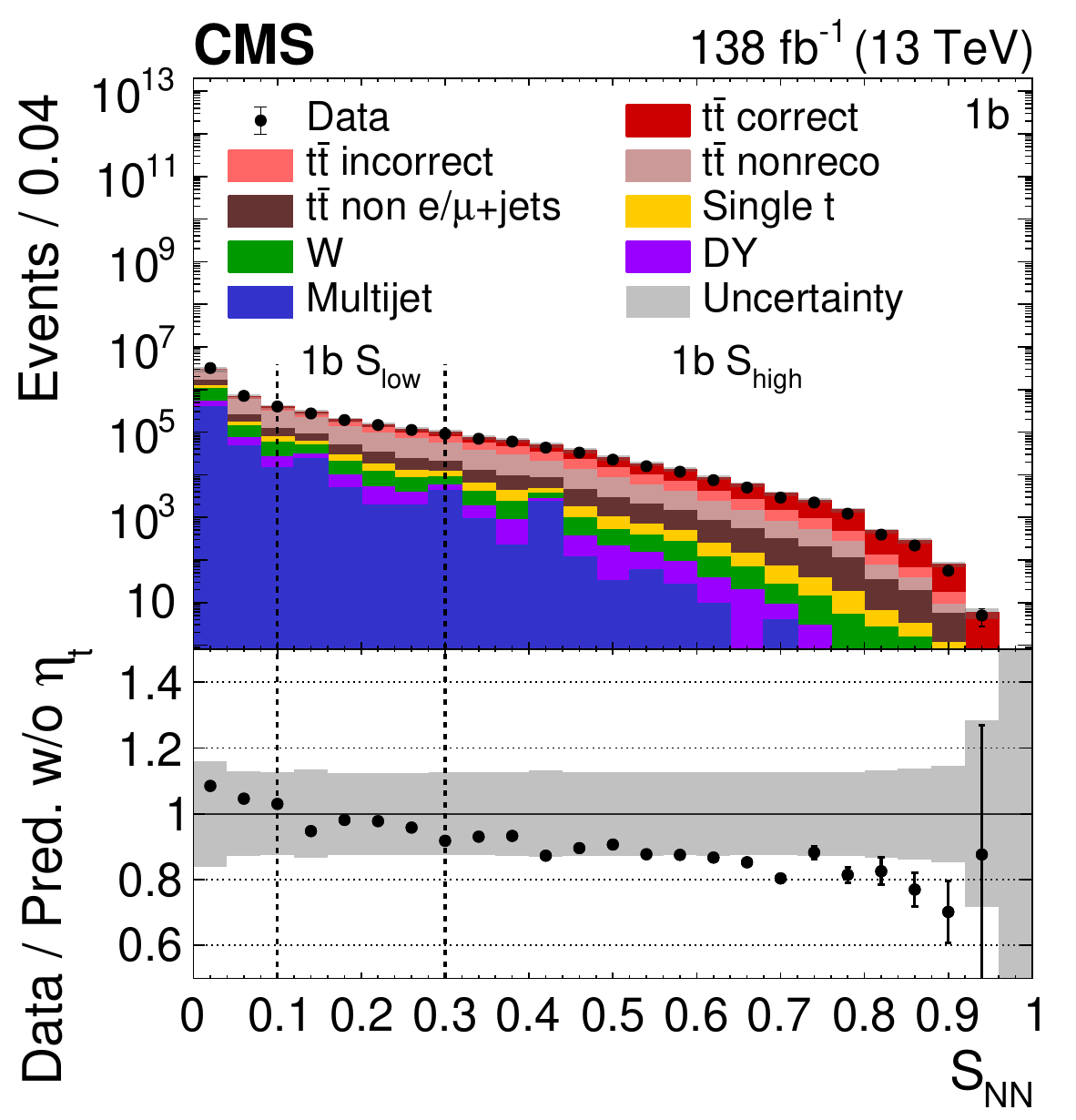}%
\hfill%
\includegraphics[width=0.48\textwidth]{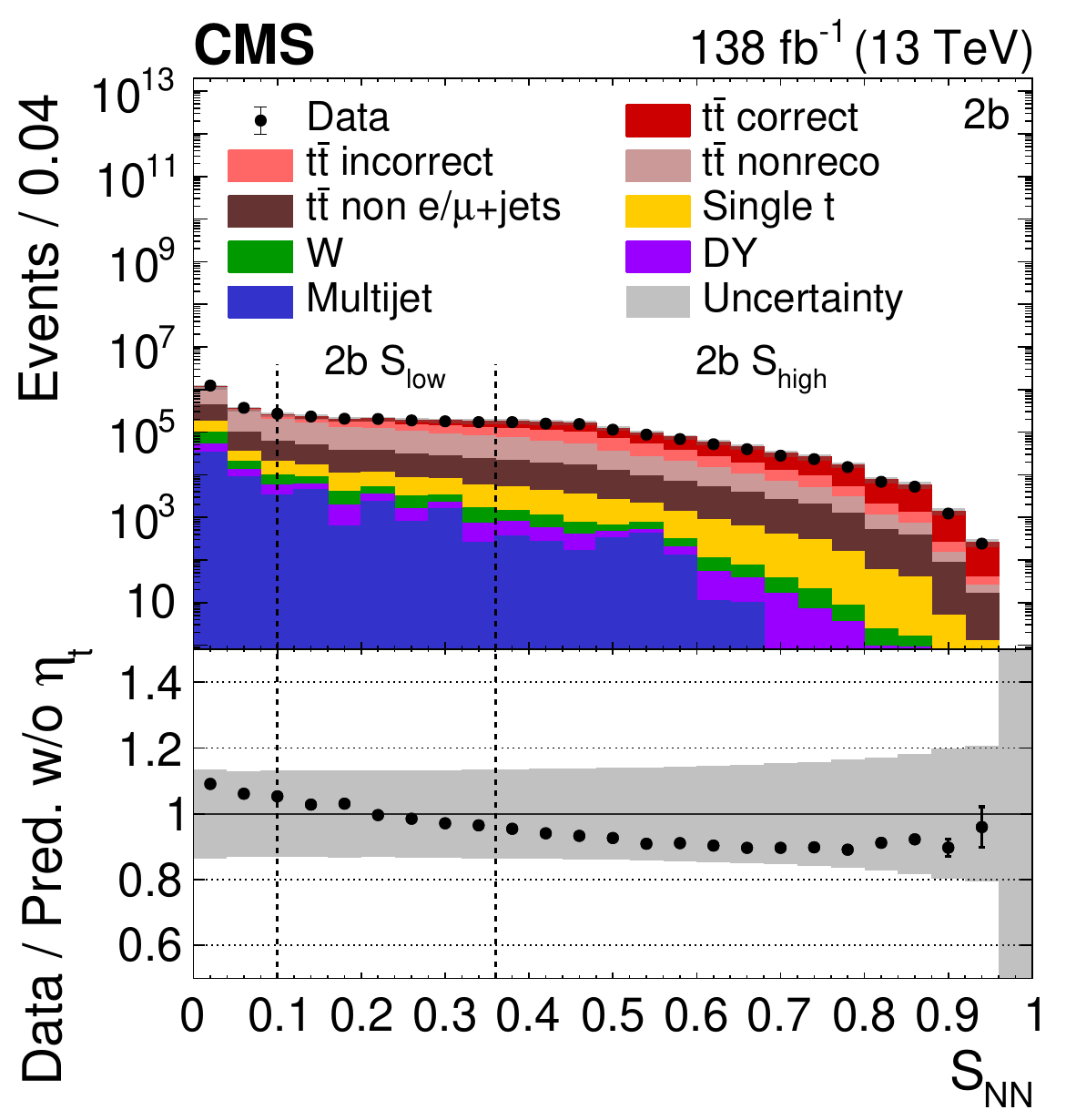}
\caption{The \SNN distribution before the mass window selections is presented for the 1\PQb (\cmsLeft) and 2\PQb (\cmsRight) categories, with data points compared against the predictions without \etat (stacked histograms). The \ttbar component is categorized into correctly reconstructed, incorrectly reconstructed, ``nonreconstructible'', and non \lpj events. The gray band represents the total statistical and systematic uncertainties in the prediction, while the vertical error bars on the data points indicate the statistical uncertainties of the data. The lower panels illustrate the ratio of observed data to predicted yields excluding \etat.}
\label{fig:reconstruction_1}
\end{figure}

Figure~\ref{fig:reconstruction_1} shows the \SNN distributions for the 1\PQb and 2\PQb categories, defined by requiring either exactly one or both of the \PQb jets from the \ttbar decay to be medium \PQb tagged. A downward trend is observed in the data relative to the expectation; however, these discrepancies are resolved by the likelihood fit described in Section~\ref{sec:signalfit}, as shown in Fig.~\ref{fig:nn_postfit}. Four signal categories are defined by subdividing the 1\PQb and 2\PQb categories based on the \SNN values. Events in the 1\PQb (2\PQb) category are assigned to \Shigh if $\SNN>0.30$ (0.36), while those that do not meet this requirement fall into the \Slow category. The \Shigh requirement provides channels with a higher fraction of correctly reconstructed events, improving the resolution of angular correlation variables. These thresholds are optimized to maximize the expected significance of the \etat signal.

\begin{figure*}[!t]
\centering
\includegraphics[width=0.495\textwidth]{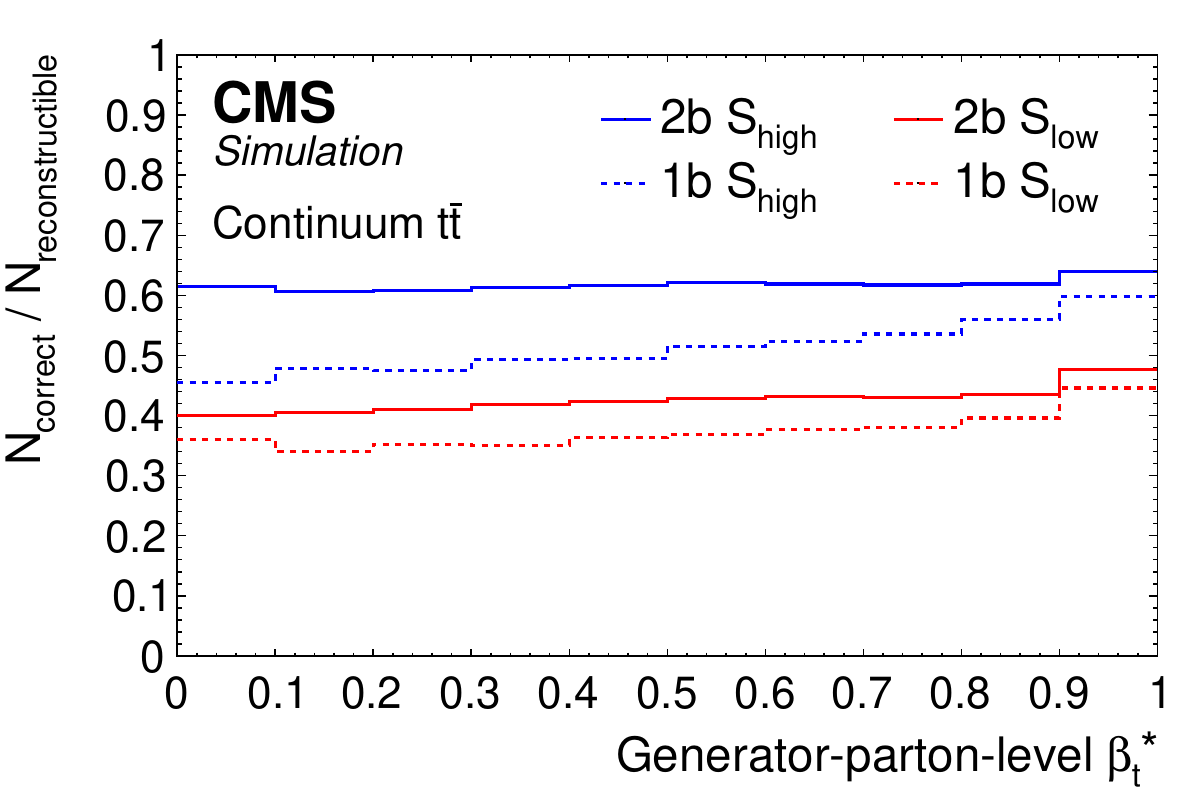}%
\hfill%
\includegraphics[width=0.495\textwidth]{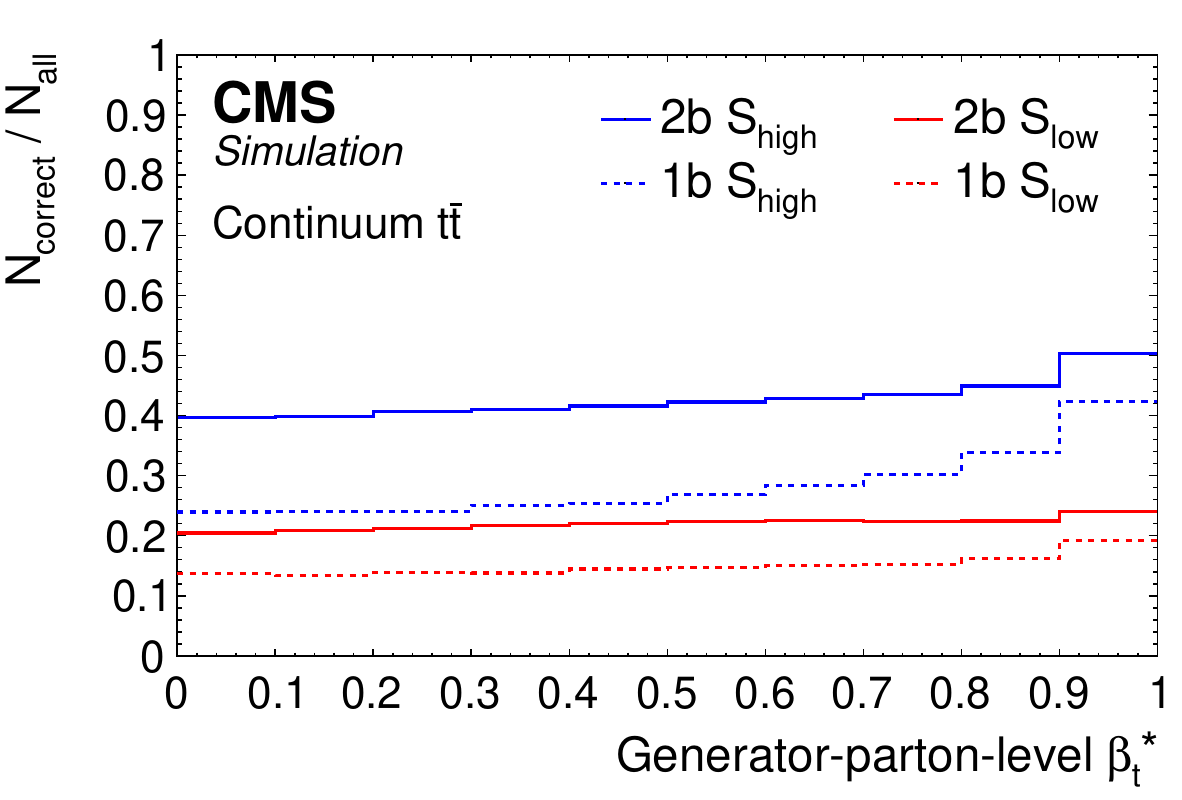}\\
\includegraphics[width=0.495\textwidth]{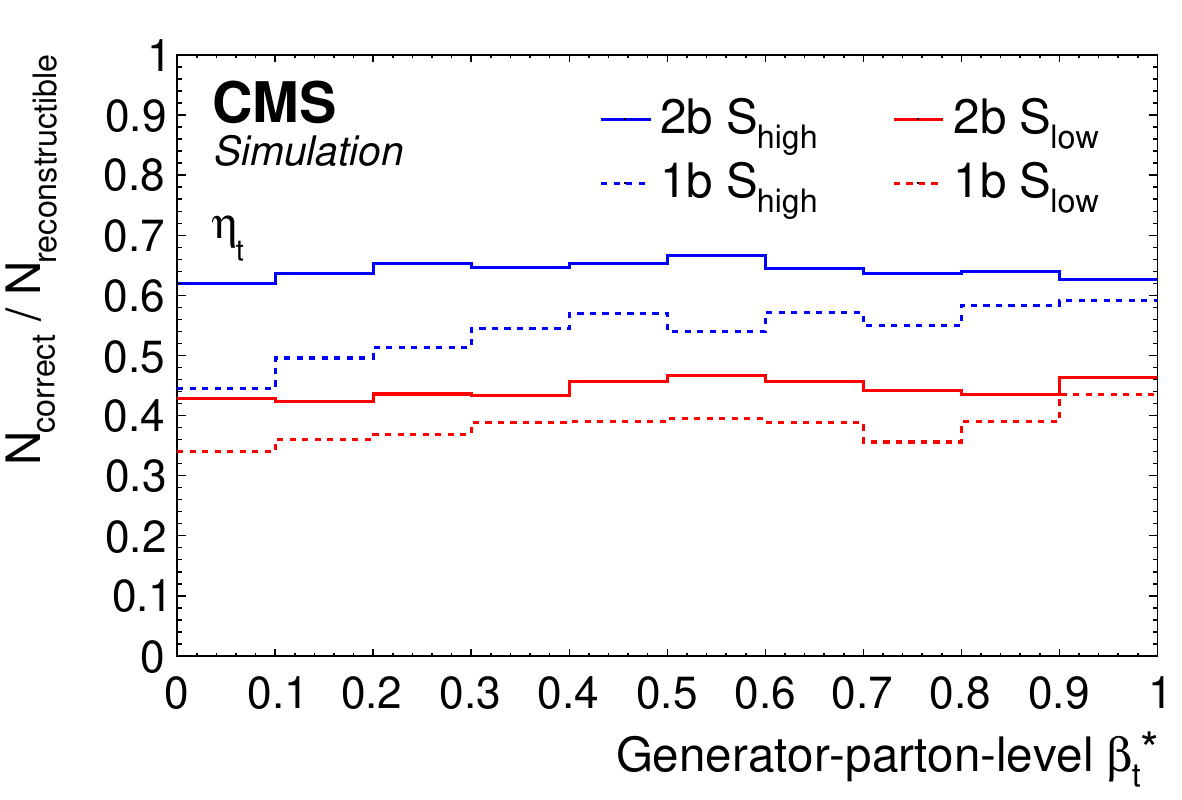}%
\hfill%
\includegraphics[width=0.495\textwidth]{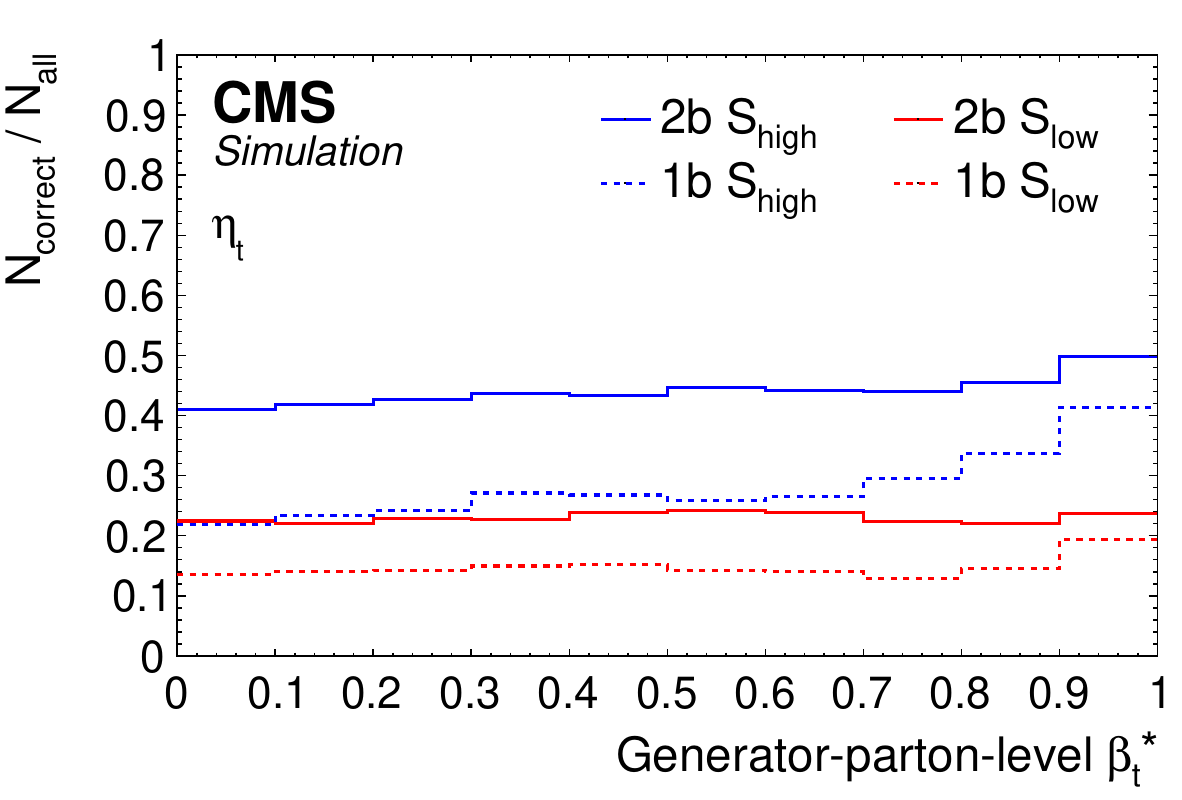}
\caption{Reconstruction performance as a function of \boost at the parton level. The plots in the left column show the fraction of correctly reconstructed events ($N_{\text{correct}}$) relative to the reconstructible events ($N_{\text{reconstructible}}$), while the plots in the right column display those relative to all generated events for each respective process ($N_{\text{all}}$). The plots in the upper row are based on the continuum \ttbar simulation, while those in the lower row use the \etat simulation. Results are presented separately for the 1\PQb and 2\PQb categories under the \Slow and \Shigh selections.}
\label{reconstruction_fig_2}
\end{figure*}

In the continuum \ttbar (\etat) simulation, the fractions of reconstructible \lpj events are 71 (63)\%, 45 (46)\%, 63 (48)\%, and 37 (34)\% for the 2\PQb~\Shigh, 2\PQb~\Slow, 1\PQb~\Shigh, and 1\PQb~\Slow categories, respectively. The generally lower fractions in the \etat sample can be explained by the fact that at low \ttm, the decay products of the top quarks tend to be softer, making them less likely to pass the \pt thresholds described in Section~\ref{sec:objects}. Figure~\ref{reconstruction_fig_2} shows the reconstruction efficiencies for the continuum \ttbar (\etat) simulation as a function of the parton-level \boost. The efficiencies are shown relative to both the reconstructible events and to the inclusive continuum \ttbar (\etat) samples. In the \etat sample, the NN correctly assigns the jets to the up- and down-type quarks in 72\% of the events, provided that the two jets from the hadronic \PW boson decay are correctly selected. The NN achieves this by utilizing the kinematic observables of the \ttbar decay products, such as the tendency for the down-type jet to be less energetic than the up-type jet~\cite{Brandenburg:2002xr}, as well as the \PQb tagging information to identify charm jets~\cite{Bols_2020, BTV-16-002}.

\section{Non-\texorpdfstring{\ttbar}{tt} background estimation}
\label{sec:background}

In addition to the dominant continuum \ttbar background, other key processes contributing to the selected events include multijet, DY, \PW boson, and single top quark production.

The expected contributions of single top quark production are 4.2, 2.4, 2.7, and 1.5\% in the 1\PQb \Slow, 1\PQb \Shigh, 2\PQb \Slow, and 2\PQb \Shigh categories, respectively. Templates representing the SM expectation for this process are derived from simulation. We incorporate relevant systematic uncertainties as described in Section~\ref{sec:uncertainties} into the fit.

The contributions from EW processes, predominantly DY and \PW boson production, as well as those from multijet production (multijet/EW) suffer from limited simulation sample sizes and are difficult to model reliably in the SR because of the large production cross sections in combination with very low selection efficiencies. Therefore, a data-driven (DD) method is employed to estimate the combined multijet/EW shape, while the normalization is determined via a simulation-based approach; both procedures are detailed in the following paragraphs. According to the simulation, these backgrounds contribute 10.0, 3.7, 2.2, and 0.5\% in the 1\PQb \Slow, 1\PQb \Shigh, 2\PQb \Slow, and 2\PQb \Shigh categories, respectively.

\begin{figure*}[!htp]
\centering
 \includegraphics[width=0.495\textwidth]{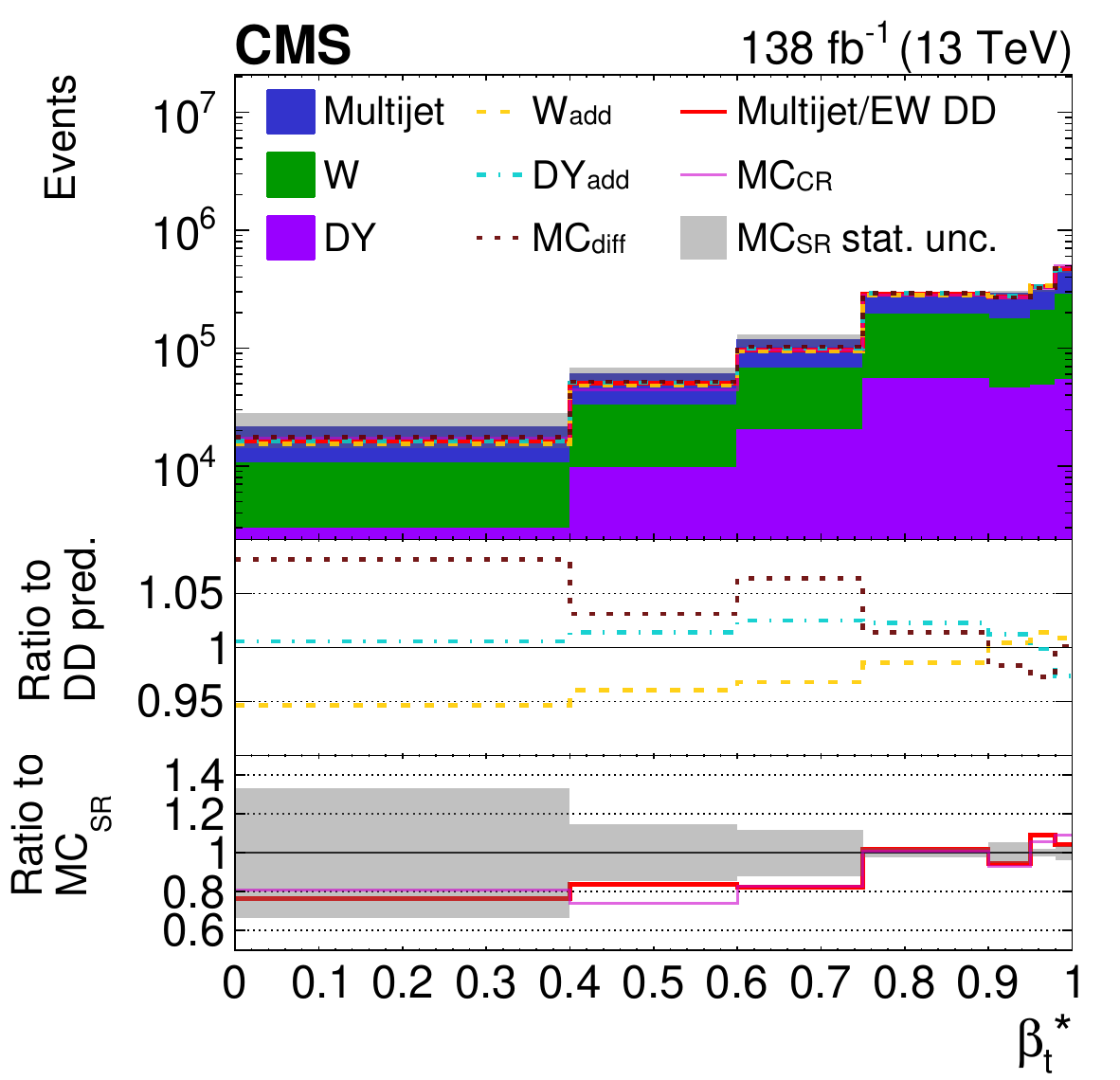}%
 \hfill%
 \includegraphics[width=0.495\textwidth]{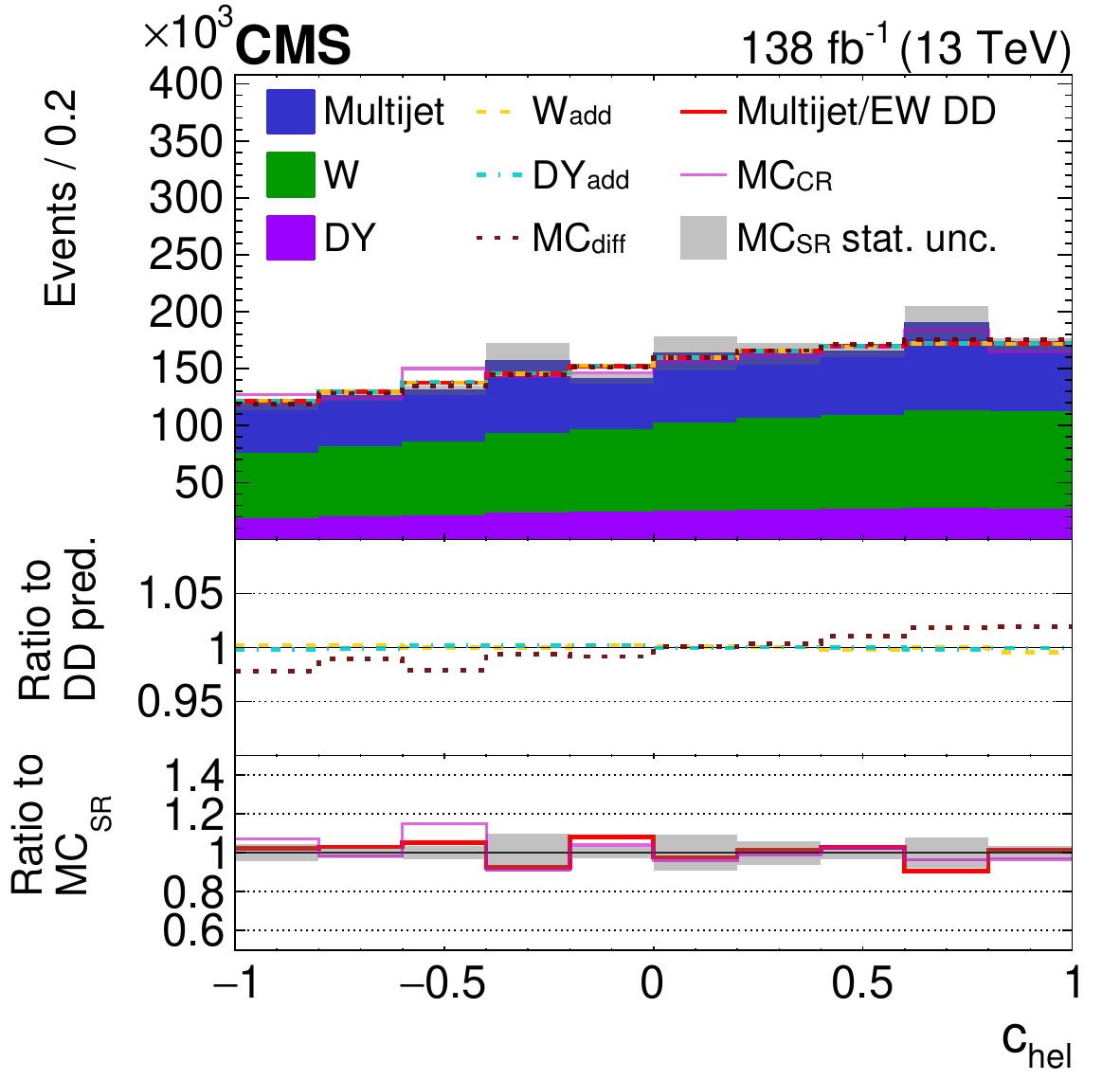} \\
 \includegraphics[width=0.495\textwidth]{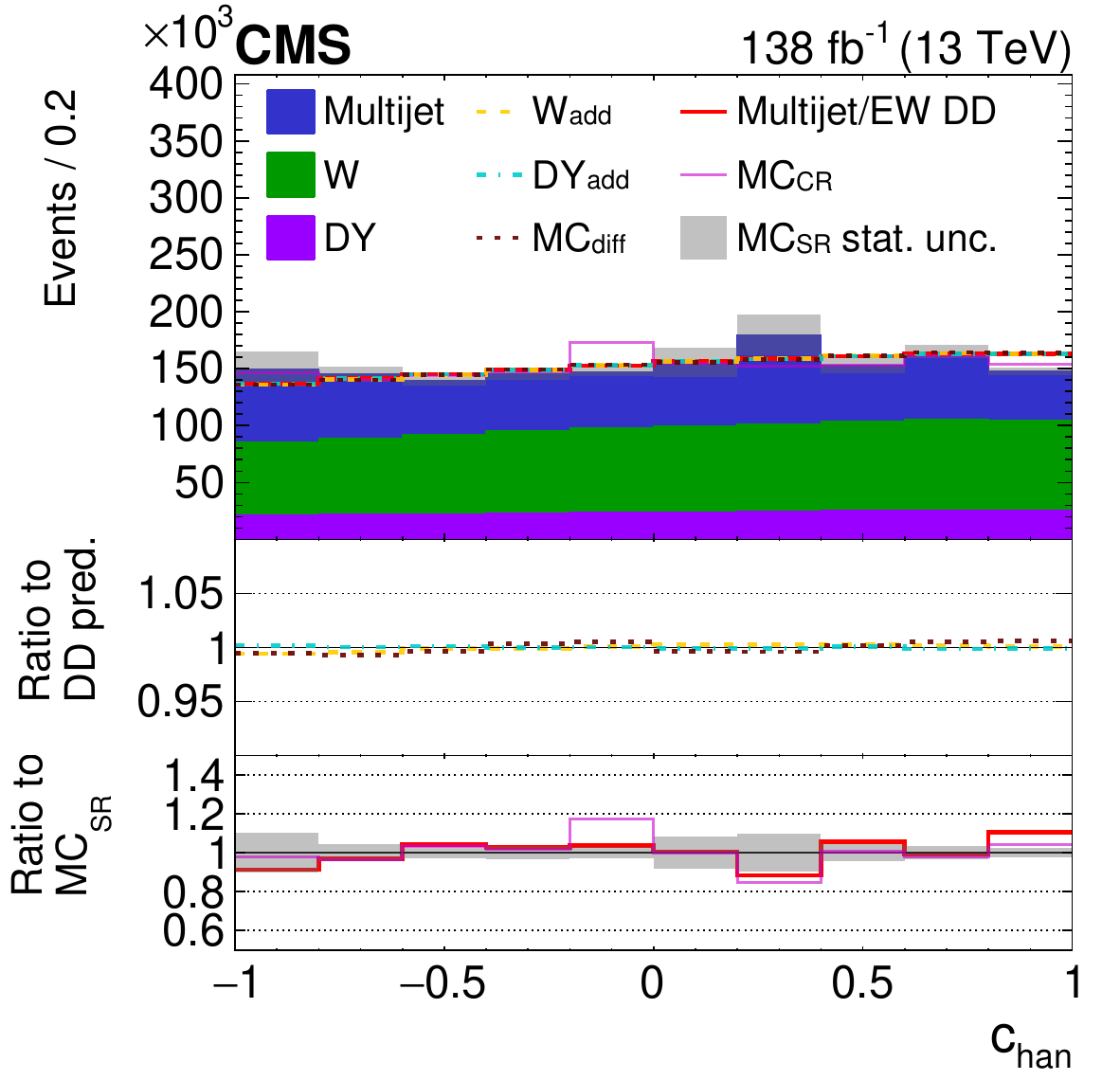}
\caption{Comparison of multijet/EW \boost (upper, left), \chel (upper, right), and \chan (lower) distributions from different sources. The upper panels show the distributions for the DD prediction (Multijet/EW DD, red line), the CR simulation ($\text{MC}_{\text{CR}}$, pink line), and the SR simulation (stacked histogram). The gray shaded band represents the statistical uncertainty of the SR simulation. Systematic uncertainties in the DD prediction are shown as dotted lines for \Wadd (yellow), \DYadd (turquoise), and \MCdiff (dark red). All distributions are normalized to the simulated SR event yield. The middle panels display the ratios of these systematic variations to the DD prediction, while the lower panels show the ratios of the DD prediction and CR simulation to the SR simulation. To reduce statistical fluctuations, the \SNN and invariant mass requirements are not applied.}
 \label{fig:background_1}
\end{figure*}

To construct the combined multijet/EW template shape, we first select events that have no jets fulfilling the medium \PQb tagging requirement. This selection defines the control region (CR) and serves to minimize contributions from \ttbar and single top quark production. However, feeding these \PQb-jet-vetoed events directly into the NN (Section~\ref{sec:reconstruction}) yields \SNN scores that are systematically too low, as the NN input also contains \PQb tagging information. To ensure compatibility with the event reconstruction procedure used in the SR, for each event in the CR, exactly one (two) jet(s) are randomly labeled as medium \PQb tagged when targeting the 1\PQb (2\PQb) signal categories. We then apply the same NN reconstruction, \SNN selections depending on the corresponding signal category, and invariant mass requirements for \tql, \tqh, and the \PW boson as detailed in Section~\ref{sec:reconstruction}. The combined multijet/EW template shape is then obtained by subtracting the simulated \ttbar and single top quark contributions from the CR data. The signal contribution in this region is negligible and is therefore not subtracted. The templates are obtained separately for electron and muon events to account for lepton-specific shape differences. We combine the data-taking periods for the template construction to reduce statistical fluctuations. As demonstrated in the lower panels of Fig.~\ref{fig:background_1}, the kinematic distributions from the DD prediction, the CR simulation, and the SR simulation are generally consistent within uncertainties.

To account for potential mismodeling and variations in the background composition, the following shape uncertainties are applied.
\begin{enumerate}
  \item {Process composition} (\Wadd, \DYadd){:} To allow for variations in the relative fractions of \PW boson, DY, and multijet production, we create two alternative versions of the combined template. These are constructed by separately adding pure simulated \PW boson and DY templates from the CR to the nominal combined template, thereby increasing their respective relative contributions. The resulting deviations from the nominal template are symmetrized and taken as systematic uncertainties to account for potential over- or underestimation of the \PW boson and DY fractions. These uncertainties are estimated separately in each signal category and lepton flavor, while combining the data-taking periods, across which they are considered correlated.
  \item {CR-to-SR transfer} (\MCdiff){:} This uncertainty accounts for template shape differences between the CR and SR. It is derived by comparing the simulated EW shape in the CR to the SR without the \SNN and invariant mass requirements to reduce statistical fluctuations. The multijet simulation is not used because the limited sample size results in unreliable shapes. This uncertainty is estimated separately for the 1\PQb and 2\PQb categories, as well as for the electron and muon templates, while combining the data-taking periods, among which they are considered correlated.
  \item {Top quark subtraction:} For every variation of the \ttbar and single top quark templates described in Section~\ref{sec:uncertainties}, a corresponding multijet/EW template is derived by subtracting the modified templates from the CR data. These variations are treated as fully correlated with the corresponding \ttbar and single top quark uncertainties in the fit.
\end{enumerate}

The normalizations of the combined background templates are derived in the SR separately for each signal category, using a simulation-based scaling method designed to overcome the lack of statistical precision in the multijet simulation. First, we estimate the number of expected background events from the simulation in the SR combined for all categories without applying the \SNN and invariant mass selections. This ensures sufficient simulated events for all processes. Afterwards, for each signal category, this combined event yield is multiplied by the fraction of events that fulfill the corresponding \SNN and invariant mass selection criteria, evaluated using only the EW simulations. The EW simulations have sufficient events to evaluate these fractions reliably. However, this procedure assumes similar selection efficiencies for the multijet and EW processes. Within the large statistical uncertainties of the multijet simulation, this assumption is compatible with the available simulated samples. Possible deviations are accounted for by assigning a 20\% normalization uncertainty to the multijet/EW background estimate, covering the maximum variation in its normalization alone needed to explain the differences between data and simulation across signal categories. This uncertainty is treated as uncorrelated across each signal category, data-taking period, and lepton flavor.

\begin{figure*}[!htp]
  \centering
  \includegraphics[width=0.495\textwidth]{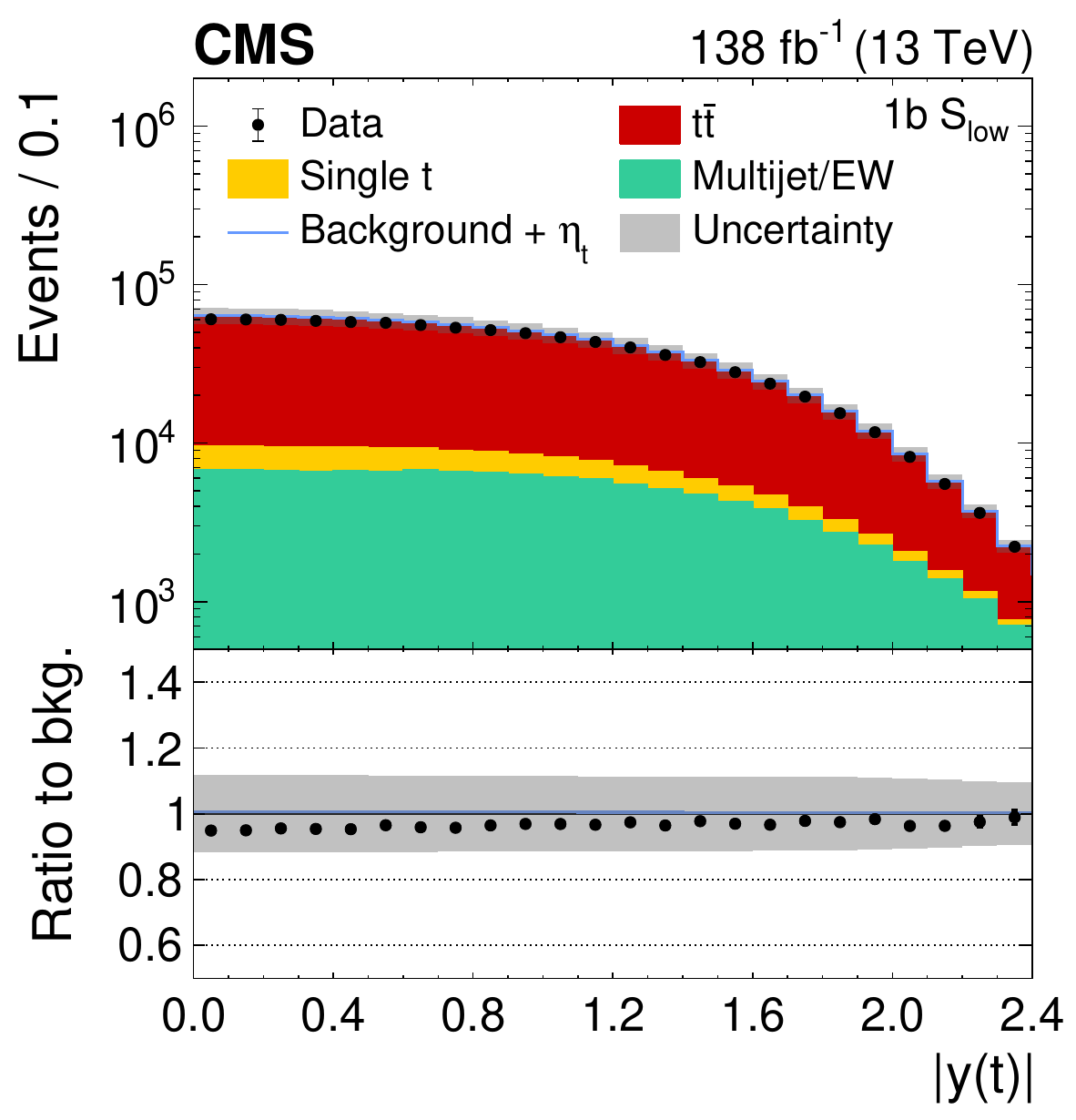}%
  \hfill%
  \includegraphics[width=0.495\textwidth]{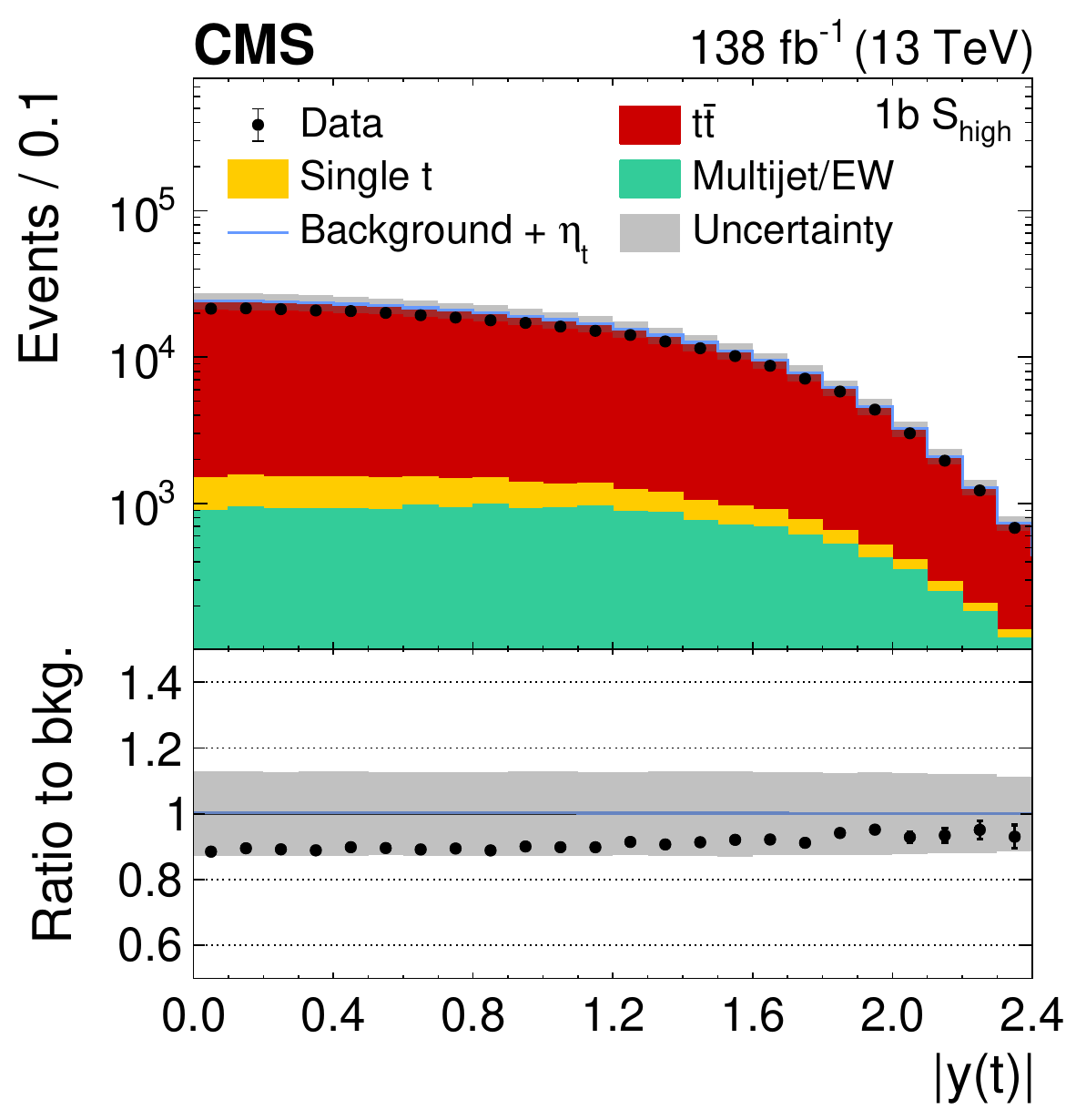} \\
  \includegraphics[width=0.495\textwidth]{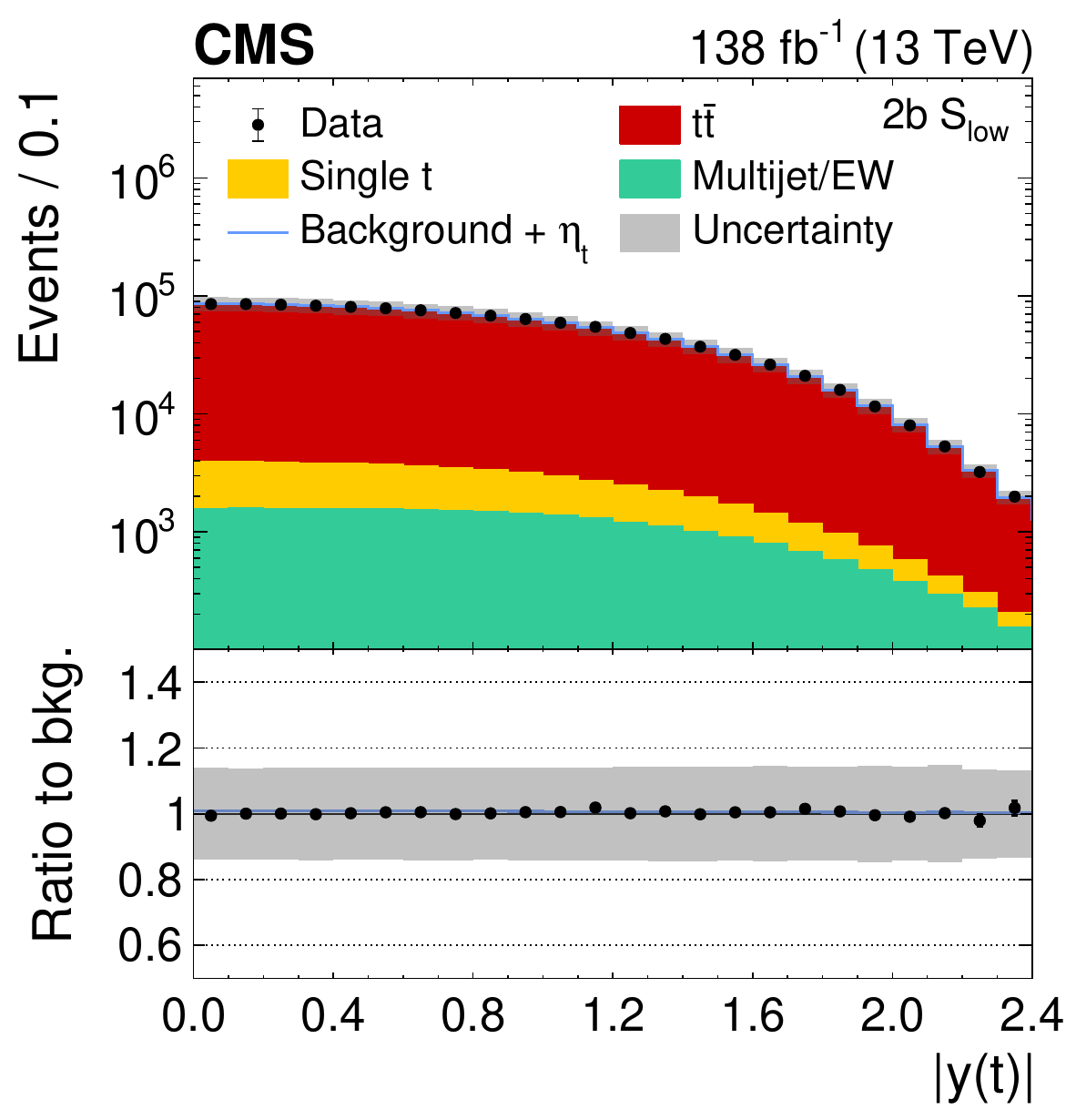}%
  \hfill%
  \includegraphics[width=0.495\textwidth]{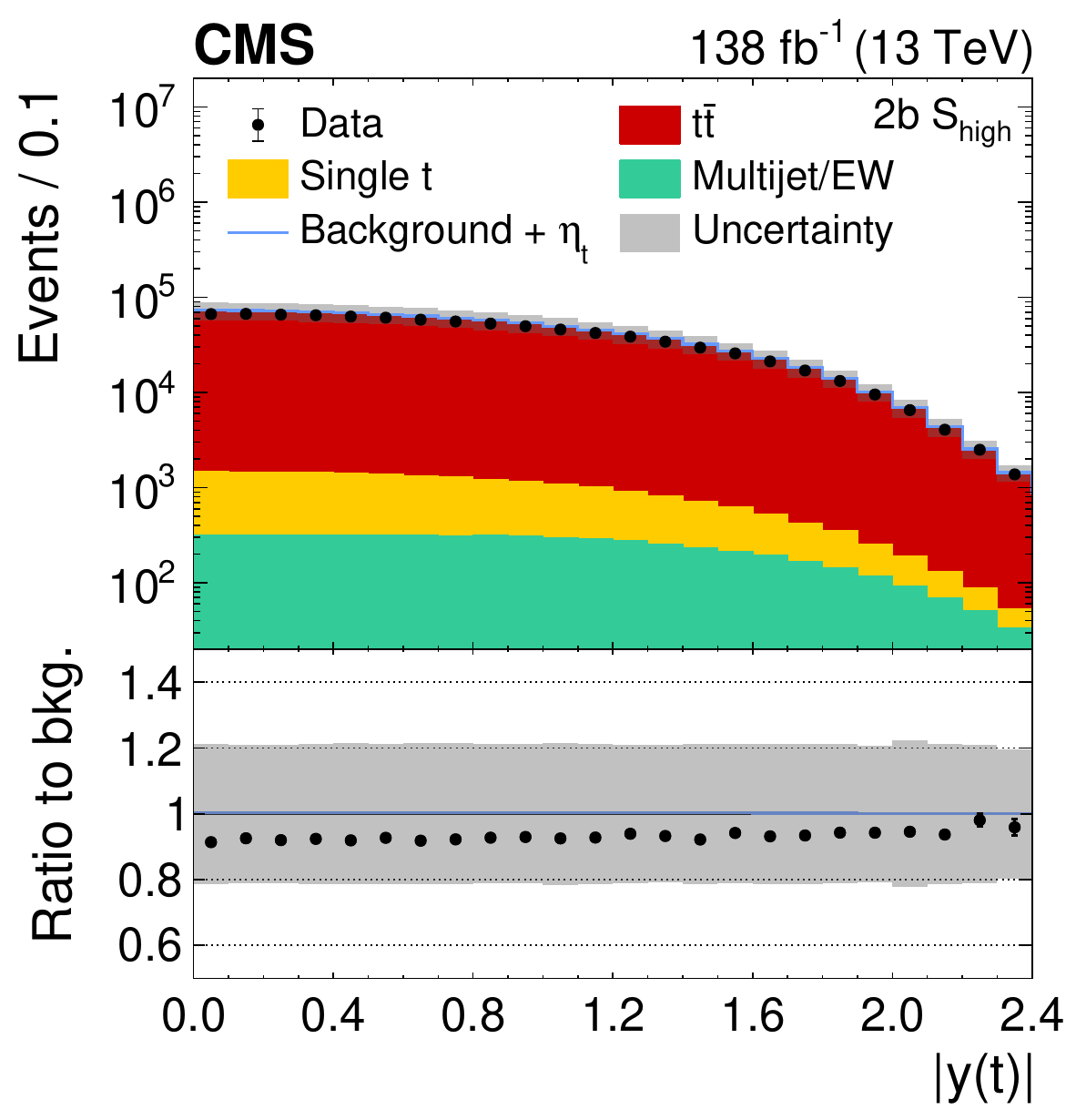}
  \caption{Pre-fit distributions of \topy in all four categories. The data (points) are compared to the predictions without \etat (stacked histograms) and those including it (blue line). The \ttbar and single top quark contributions are taken from simulation, while the multijet/EW background is derived from the CR using the DD method. The gray uncertainty band represents the combined statistical and systematic uncertainties in the predictions, while the vertical bars on the data points indicate their statistical uncertainties. The ratios to the predicted yields excluding \etat are shown in the lower panels.}
  \label{fig:background_2}
\end{figure*}

\begin{figure*}[!htp]
  \centering
  \includegraphics[width=0.495\textwidth]{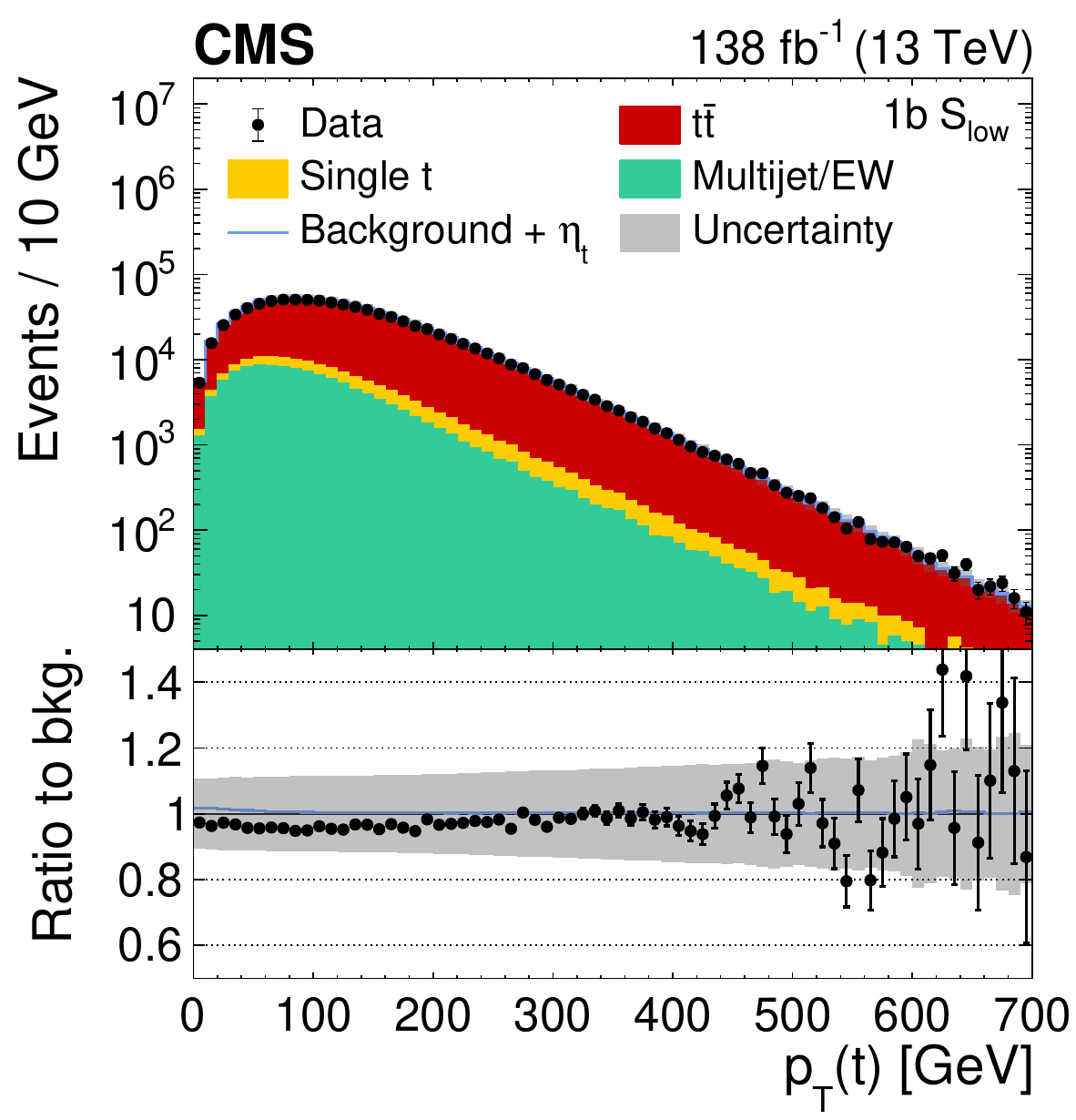}%
  \hfill%
  \includegraphics[width=0.495\textwidth]{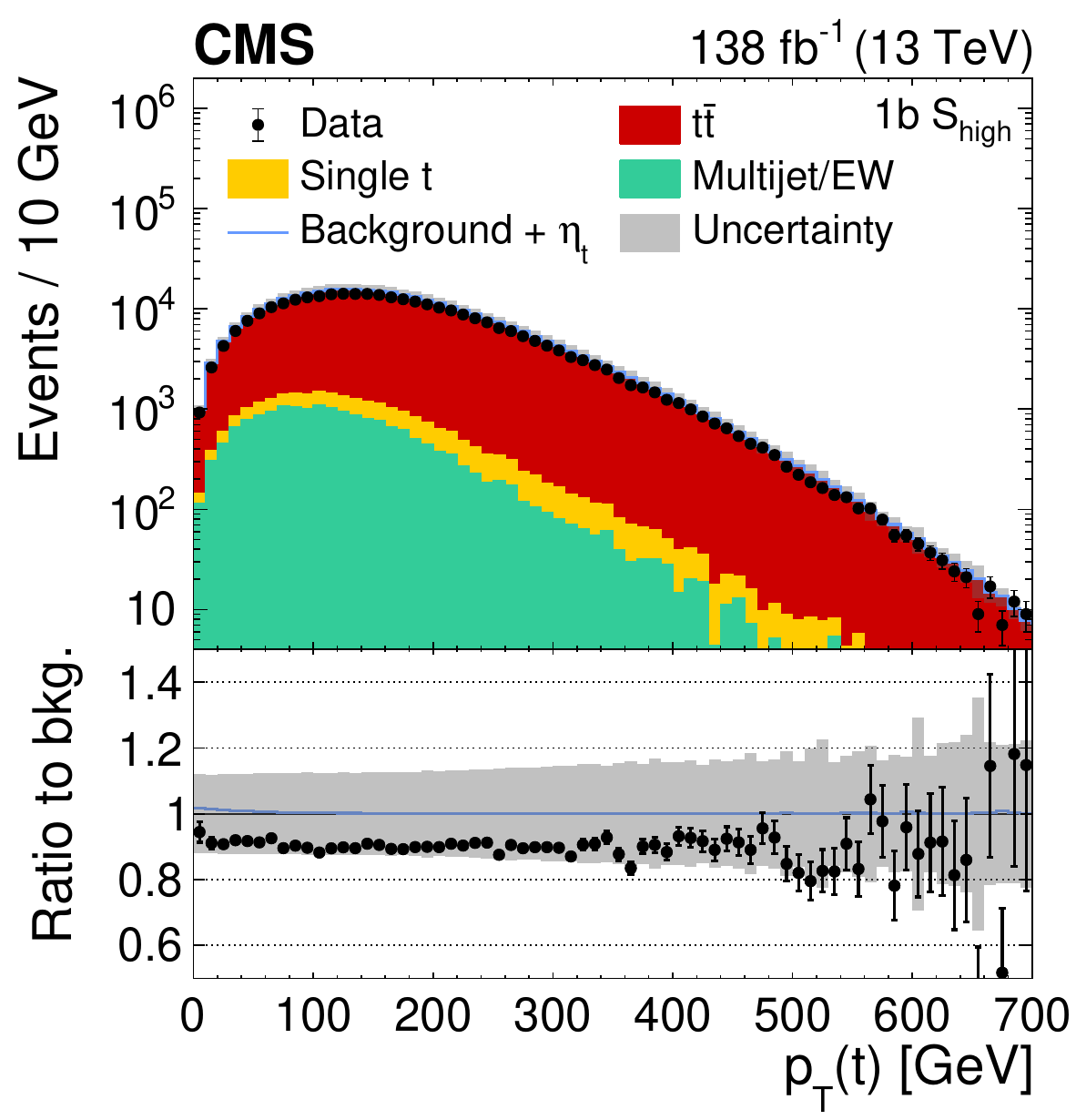} \\
  \includegraphics[width=0.495\textwidth]{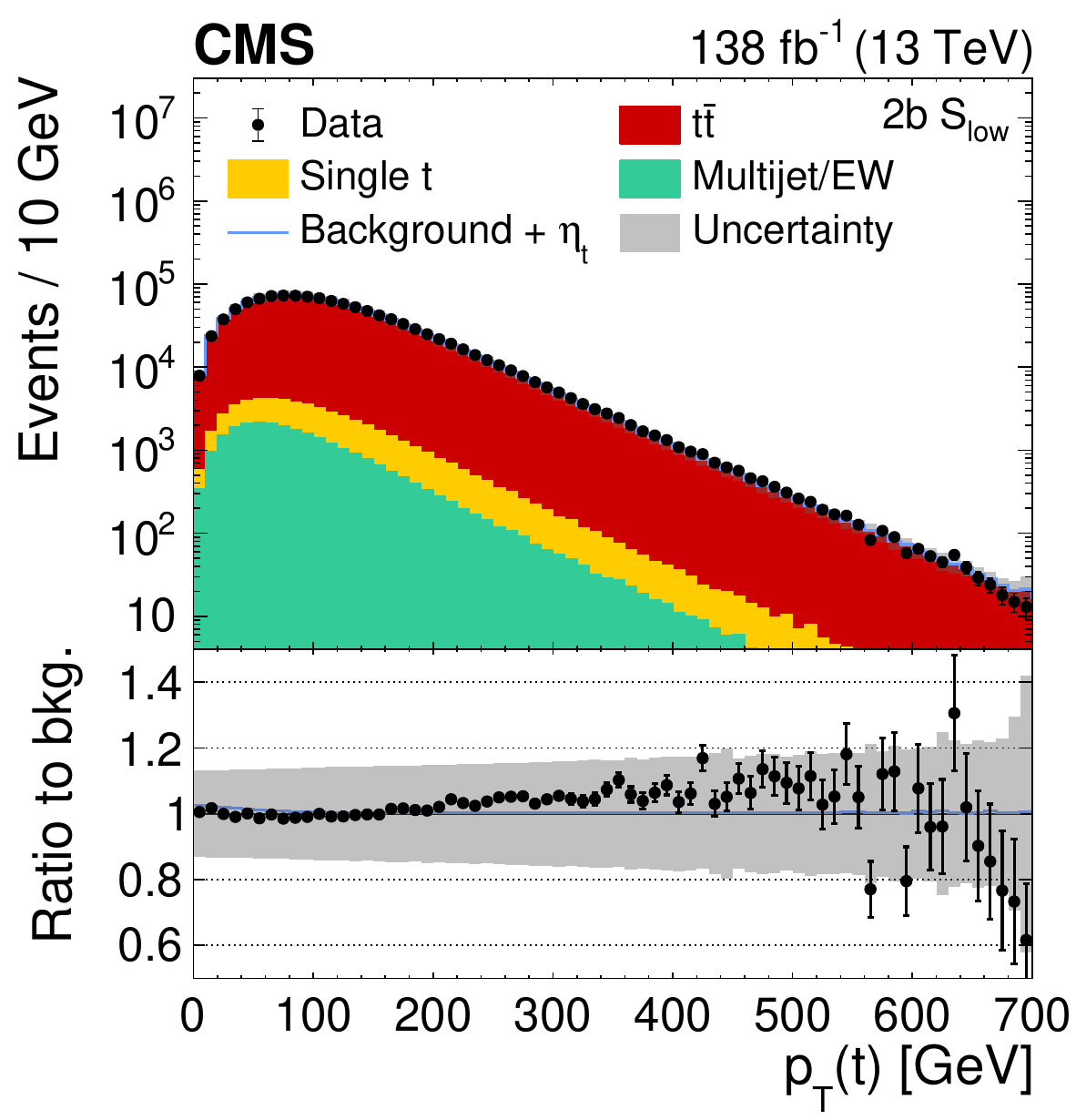}%
  \hfill%
  \includegraphics[width=0.495\textwidth]{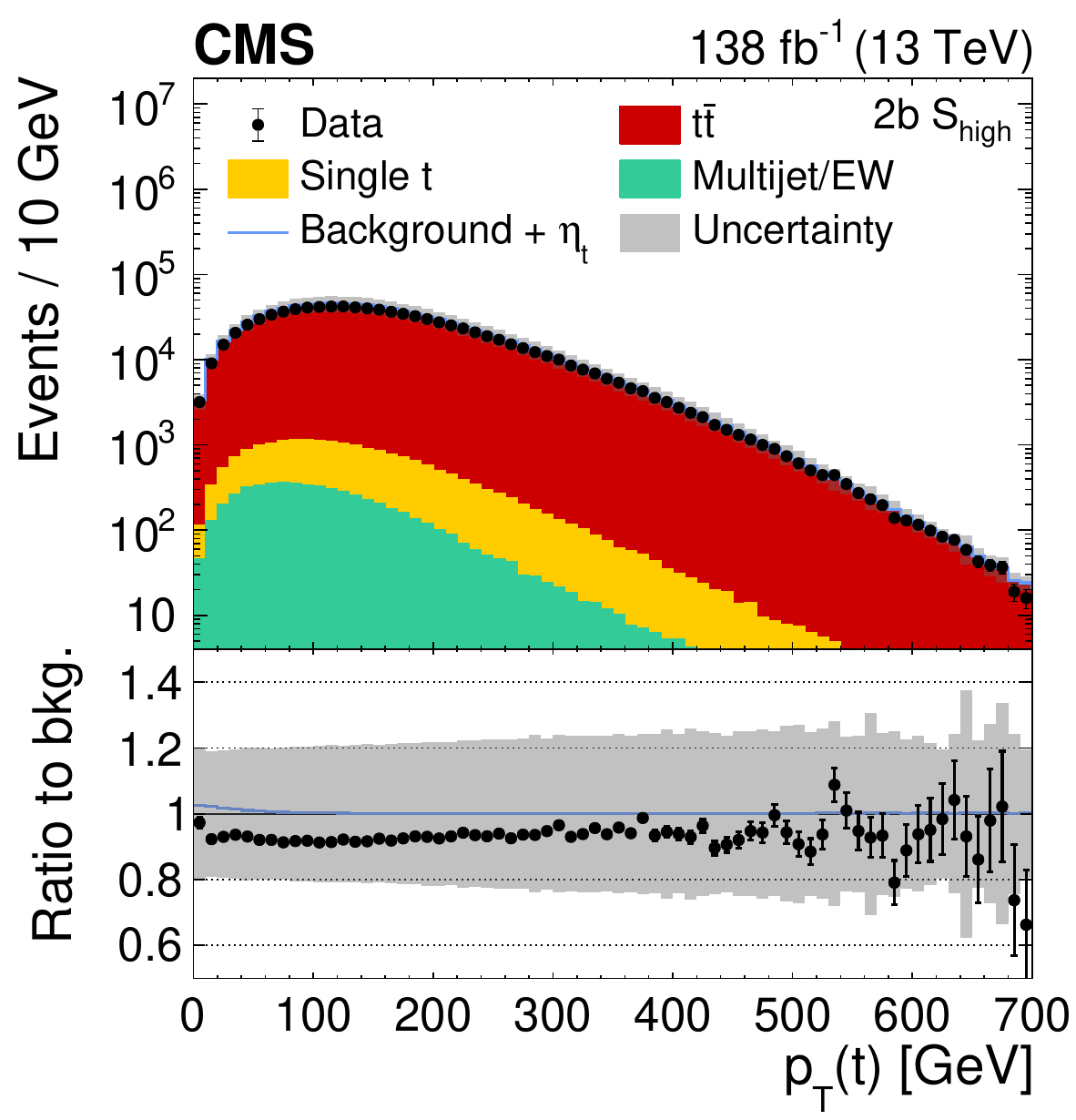}
  \caption{Pre-fit distributions of \tpt in all four categories. The data (points) are compared to the predictions without \etat (stacked histograms) and those including it (blue line). The \ttbar and single top quark contributions are taken from simulation, while the multijet/EW background is derived from the CR using the DD method. The gray uncertainty band represents the combined statistical and systematic uncertainties in the predictions, while the vertical bars on the data points indicate their statistical uncertainties. The ratios to the predicted yields excluding \etat are shown in the lower panels.}
  \label{fig:background_3}
\end{figure*}

\begin{figure*}[!htp]
  \centering
  \includegraphics[width=0.495\textwidth]{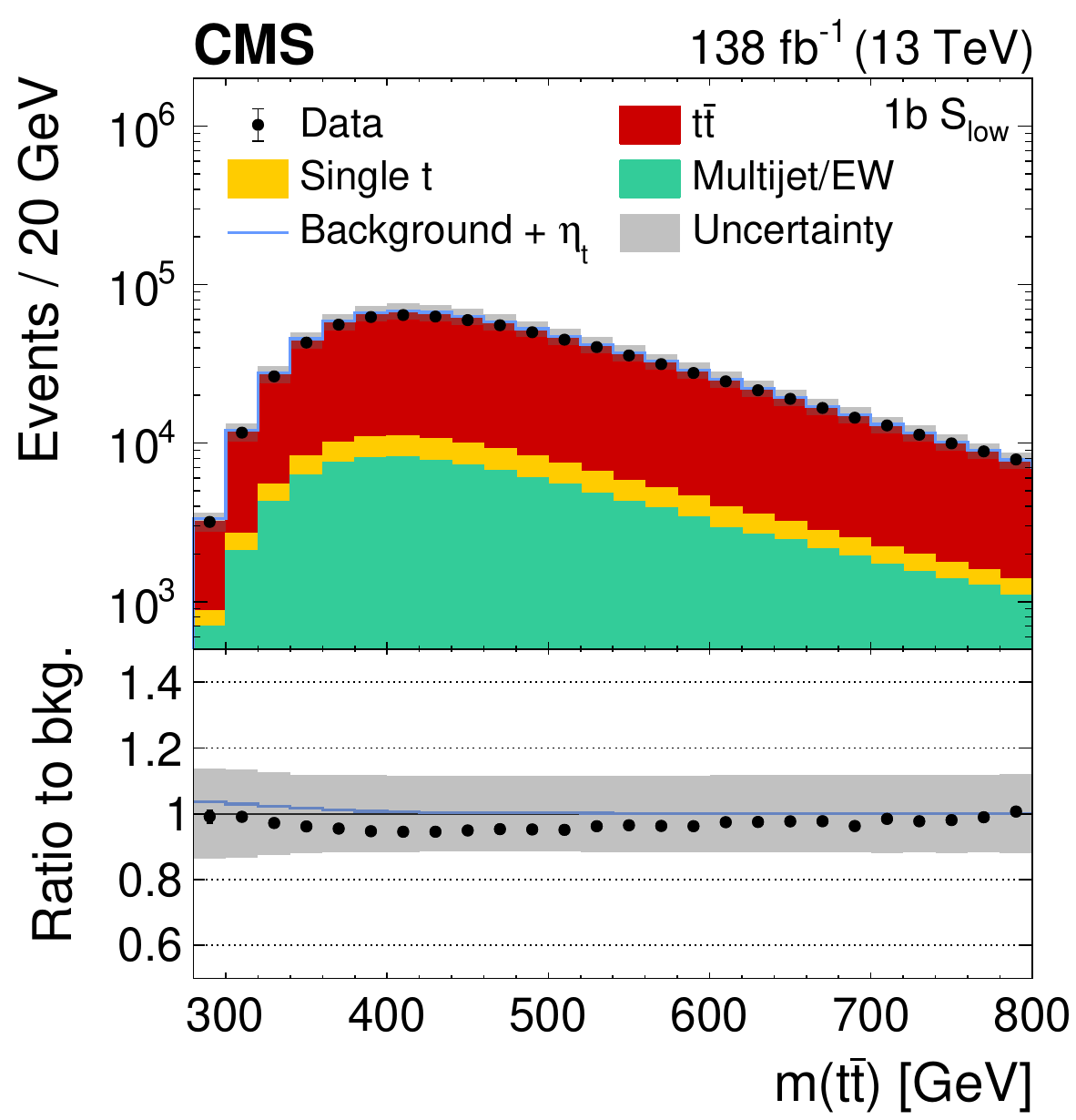}%
  \hfill%
  \includegraphics[width=0.495\textwidth]{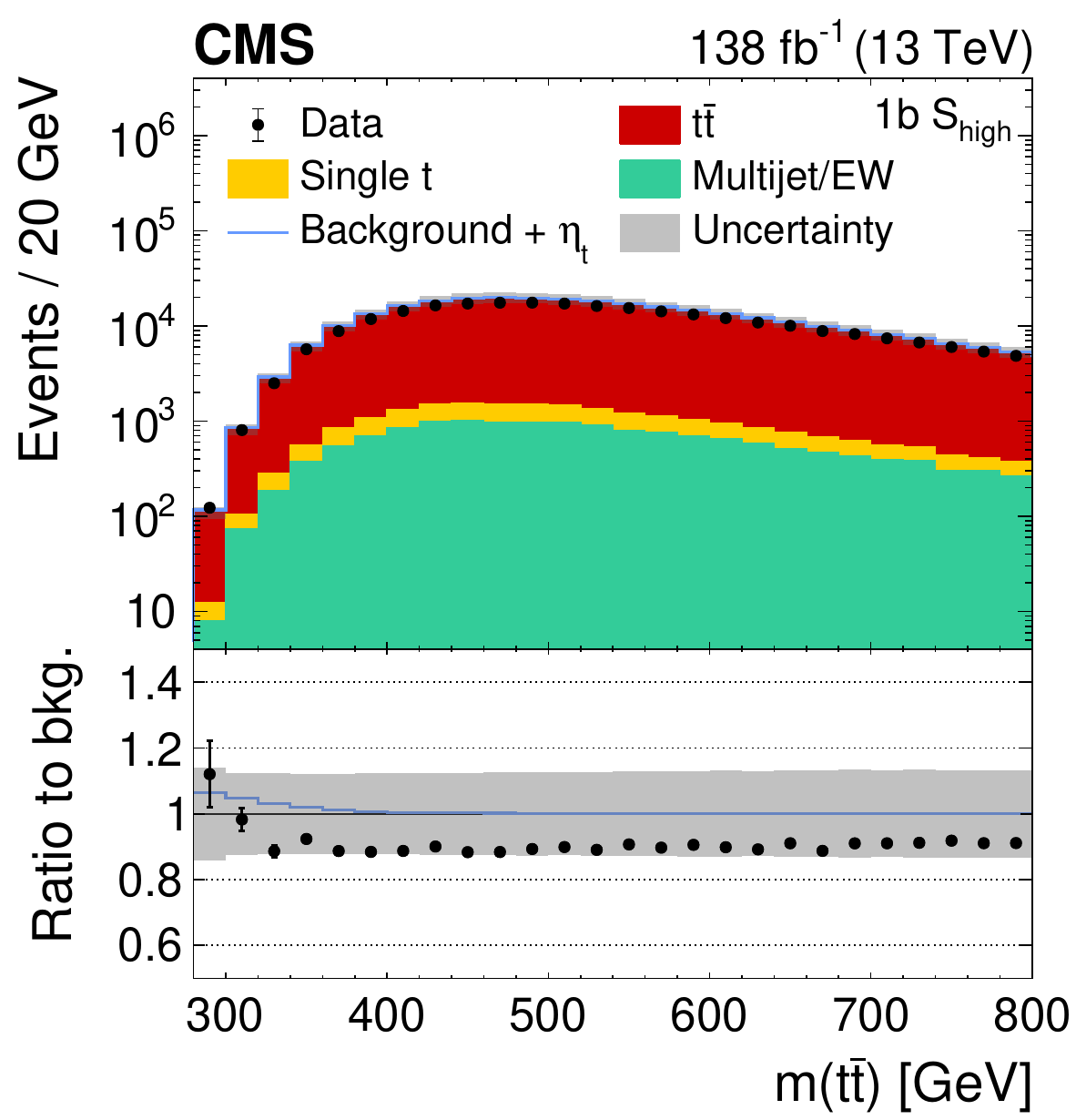} \\
  \includegraphics[width=0.495\textwidth]{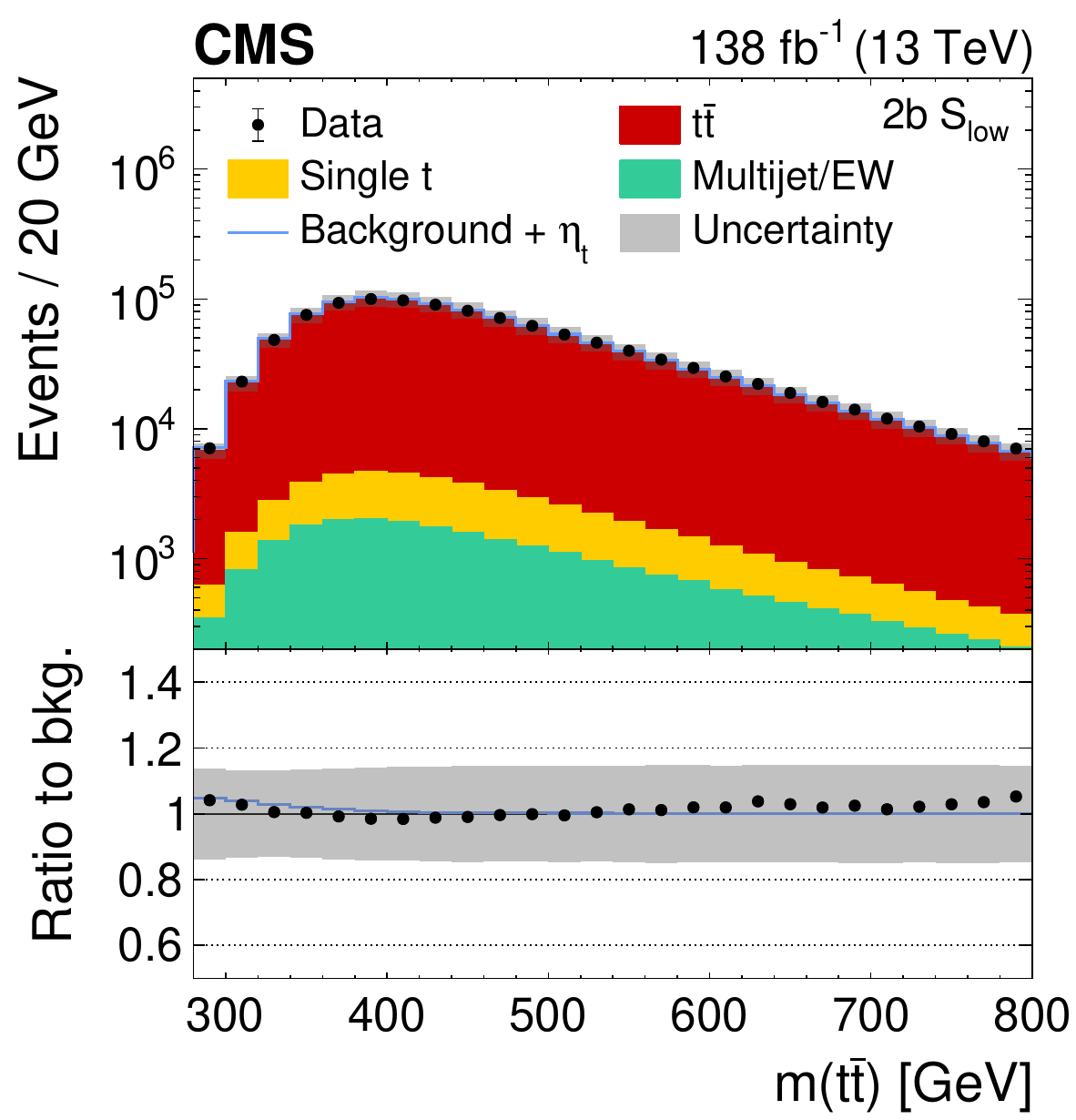}%
  \hfill%
  \includegraphics[width=0.495\textwidth]{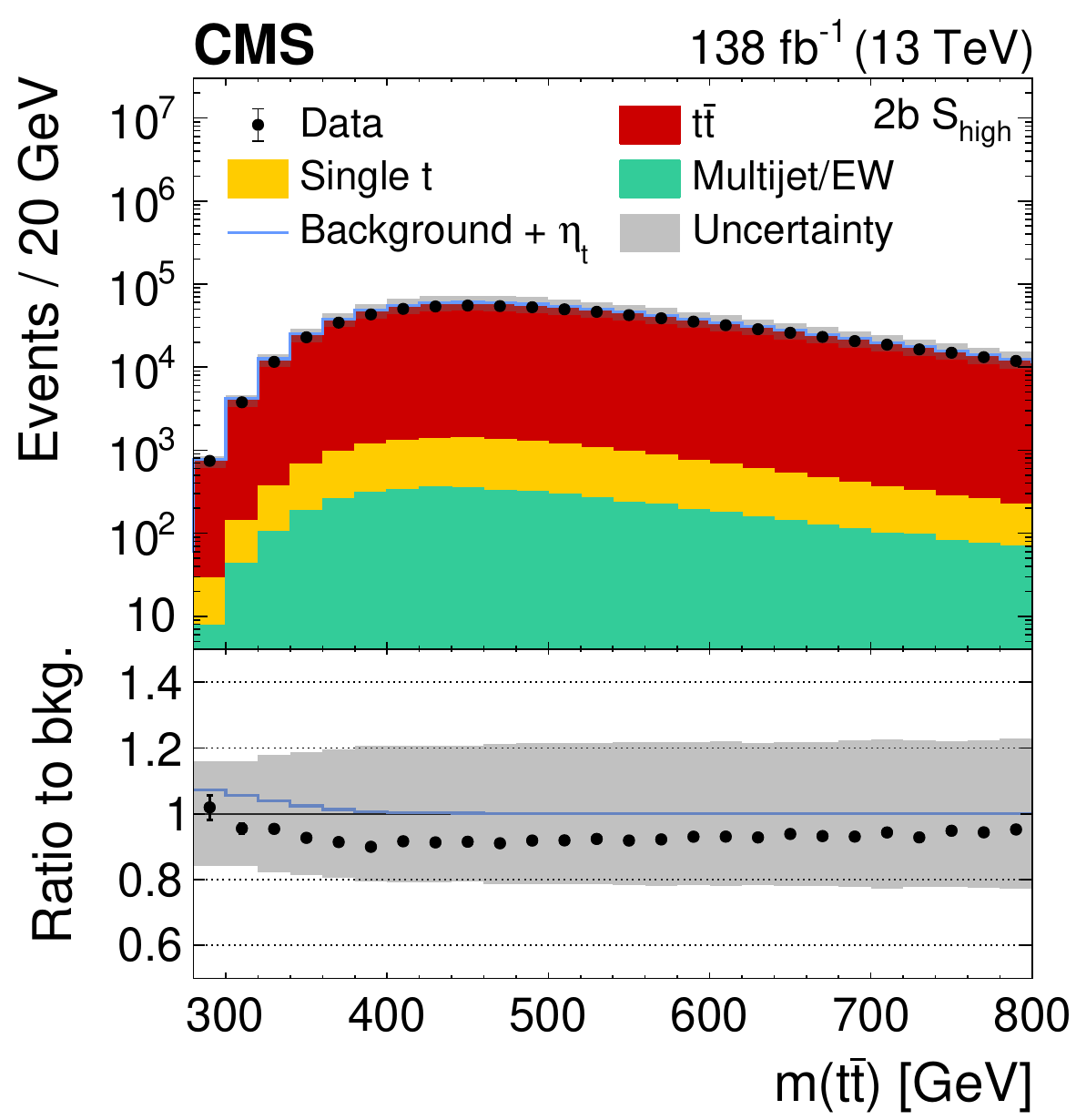}
  \caption{Pre-fit distributions of \ttm in all four categories. The data (points) are compared to the predictions without \etat (stacked histograms) and those including it (blue line). The \ttbar and single top quark contributions are taken from simulation, while the multijet/EW background is derived from the CR using the DD method. The gray uncertainty band represents the combined statistical and systematic uncertainties in the predictions, while the vertical bars on the data points indicate their statistical uncertainties. The ratios to the predicted yields excluding \etat are shown in the lower panels.}
  \label{fig:background_4}
\end{figure*}

Figures~\ref{fig:background_2}--\ref{fig:background_4} compare the observed data to the predictions, incorporating the background estimation method described in this section. The figures display several kinematic observable distributions in all signal categories to demonstrate a reasonable description of the data by the predictions. Small normalization discrepancies are observed in the \Shigh categories as a result of the downward trend of the data relative to the prediction in the \SNN distribution shown in Fig.~\ref{fig:reconstruction_1}. After the fit described in Section~\ref{sec:signalfit} corrects this trend as shown in Fig.~\ref{fig:nn_postfit}, this issue is resolved.

\section{Signal extraction}
\label{sec:signalfit}

Data are counted in 63 exclusive bins based on \boost, \chel, and \chan. These bins provide different sensitivities to the signal and background processes. In detail, they are defined by binning \boost according to $[0, 0.4, 0.6, 0.75, 0.9, 0.95, 0.98, 1]$, while both \chel and \chan use edges at $[-1, -1/3, 1/3, 1]$. The edges of the first three \boost bins were optimized to maximize the expected sensitivity of \etat and ensure that enough simulated events are available to construct reliable predictions. The \chel and \chan edges were chosen to be consistent with Ref.~\cite{TOP-24-007}. In the following, the event yields in these bins are presented as a one-dimensional histogram, where the bin index $n$ is given by $n = 9n_0+3n_1+n_2$, with $n_0$, $n_1$, and $n_2$ being the bin indices for \boost, \chel, and \chan, respectively.

A binned profile maximum likelihood fit is performed using this histogram, combining information from the four selection categories (2\PQb \Shigh, 2\PQb \Slow, 1\PQb \Shigh, 1\PQb \Slow) across four data-taking periods. The expected number of events in each bin is modeled as:
\begin{equation}
    N_{mn}\big(\mu,\{\nu_k\}\big) = \mu\,S_{mn}\big(\{\nu_k\}\big)+B_{mn}\big(\{\nu_k\}\big),
\end{equation}
where $m$ represents the category. The terms $S_{mn}$ and $B_{mn}$ correspond to the number of events expected from \etat and the background, respectively. Both terms depend on nuisance parameters $\{\nu_k\}$, which model variations in expected event yields due to the systematic uncertainties described in Section~\ref{sec:uncertainties}. The signal strength $\mu$ scales the expected signal yield.

Finally, the function
\ifthenelse{\boolean{cms@external}}{\begin{equation}\begin{aligned}
    -2\log\mathcal{L}\big(\mu,\{\nu_k\}\big) &= -2\sum_{mn} \Big(d_{mn}\log N_{mn}\big(\mu,\{\nu_k\}\big) \\
    &\quad-N_{mn}\big(\mu,\{\nu_k\}\big)\Big)-2\log G\big(\{\nu_k\}\big)
\end{aligned}\end{equation}}{\begin{equation}
    -2\log\mathcal{L}\big(\mu,\{\nu_k\}\big) = -2\sum_{mn} \Big(d_{mn}\log N_{mn}\big(\mu,\{\nu_k\}\big)-N_{mn}\big(\mu,\{\nu_k\}\big)\Big)-2\log G\big(\{\nu_k\}\big)
\end{equation}}
is minimized with respect to the parameters $\mu$ and $\{\nu_k\}$, where $d_{mn}$ is the observed event yield, and $G\big(\{\nu_k\}\big)$ represents Gaussian constraints on the nuisance parameters.

\section{Systematic uncertainties}
\label{sec:uncertainties}

Various systematic uncertainties, both theoretical and experimental, impact the predicted event yields. When extracting the \etat signal strength, we incorporate these uncertainties by using templates from alternative or reweighted simulations that reflect variations in each uncertainty, typically adjusted by one standard deviation (SD).

For the theoretical uncertainties, given that they impact every data-taking period in an identical manner, the associated nuisance parameters are completely correlated across these periods. The considered theoretical uncertainties include:
\begin{itemize}
\item {Higher-order ME corrections:} The impact of higher-order corrections in the ME calculation is estimated by varying \mur and \muf up and down by a factor of two. Despite the fact that the NNLO QCD reweighting is used for the central prediction, we derive the NLO QCD scale variations from the \PWGPYT simulation by applying the corresponding event weights. In the fit, these scale changes are modeled by two separate nuisance parameters.
\item {PDF variations:} PDF uncertainties are assessed using the 100 Hessian variations of the NNPDF3.1~\cite{NNPDF31} set along with a strong coupling \alpS variation. Each variation is implemented via its own nuisance parameter, with the corresponding distribution derived via event reweighting.
\item {Parton shower:} To evaluate uncertainties in both the initial- and final-state parton showers, the scales used for the various splitting processes, namely $\Pg\to\PQq\PAQq$, $\Pg\to\Pg\Pg$, $\PQq\to\PQq\Pg$, and $\PQb\to\PQb\Pg$, are varied up and down by a factor of two. This procedure yields eight independent variations. The resulting distributions are derived via event reweighting.
\item {Parton shower modeling:} The uncertainty in the parton shower modeling is estimated using a separate continuum \ttbar sample obtained using \PWGHRW7.1~\cite{Bahr:2008pv} with tune CH3~\cite{CMS:2020dqt}. The difference between this sample and the default \PWGPYT simulation is taken as a systematic uncertainty.
\item {Matching of ME and PS (\hdamp):} In the \POWHEG simulation, the \hdamp scale divides the phase space of the first QCD emission into soft and hard parts. For the CP5 tune, its values are $1.38\,^{+0.92}_{-0.51}\,\Mtop$~\cite{CP5}. Dedicated simulation samples are produced for each \hdamp variation to derive the corresponding distributions. Additionally, to mitigate statistical fluctuations, an NN-based method~\cite{mlreweighting} using the generator-level top quarks kinematic properties is applied to reweight the central simulation to replicate the simulations with varied \hdamp values.
\item {Top quark mass:} To account for the uncertainty in \Mtop, a variation of $\pm$0.5\GeV~\cite{CMS:2015lbj} around the nominal value of $\Mtop=172.5\GeV$ is implemented using continuum \ttbar simulations with varied \Mtop values. When estimating the expected event yields, continuum \ttbar production cross sections that depend on \Mtop are used: 843\unit{pb} for $\Mtop = 172.0\GeV$ and 820\unit{pb} for $\Mtop = 173.0\GeV$~\cite{mitov}.
\item {Top quark Yukawa coupling:} The top quark Yukawa coupling in the EW corrections is varied by $\pm$0.1 as suggested by the measurement in Ref.~\cite{Hcouplings}.
\item {EW corrections:} The difference between the multiplicative approach described in Eq.~\eqref{eq:Kfac} and the additive approach, where the EW corrections are simply added to the \ttbar continuum prediction at NNLO QCD, is taken as an uncertainty in the EW corrections~\cite{evan19}.
\item {Underlying event modeling:} We estimate the uncertainty in the underlying event modeling using dedicated sets of samples, which form an envelope of the uncertainties within the \PYTHIA CP5 tune~\cite{CP5}. The NN-based method for mitigating statistical fluctuations is also applied to this uncertainty.
\item {Leptonic decay fraction of \PQb hadrons:} The fraction of \PQb hadrons that decay leptonically is adjusted via event reweighting to reflect the current uncertainty in the branching fraction~\cite{ParticleDataGroup:2024}.
\item {Color reconnection:} The uncertainties in the color reconnection are assessed using three alternative models. The first model activates the reconnection of colored particles from resonant decays in \PYTHIA, which is deactivated in the default tune. The second model employs the gluon move approach, while the third follows a QCD-inspired scheme~\cite{Argyropoulos:2014zoa, Christiansen:2015yqa}. The differences between these three models and the default samples are treated as separate uncertainties using the NN-based reweighting method.
\item {Top quark recoil:} In simulating second and subsequent gluon emissions from bottom quarks in $\PQt\to\PW\PQb$ decays, the choice of the recoiler particle is ambiguous~\cite{Brooks:2019xso}. The default \PWGPYT continuum \ttbar simulation treats the bottom quark as the recoiler for these emissions. A separate sample is generated by treating the top quark as the recoiler instead. The difference between these two samples is considered a systematic uncertainty~\cite{ATLAStm2023} and is evaluated using the NN-based reweighting method.
\item {Off-shell top quark production:} The \ttbar simulation at NNLO QCD obtained with \MINNLO does not include off-shell top quark contributions outside a generated top quark mass window of $172.5\pm13.1\GeV$. To account for the effect of removing these off-shell top quarks, an uncertainty is derived using the NLO \POWHEG simulation by applying this same kinematic selection at the parton level. The difference between the prediction with and without this selection is taken as a systematic uncertainty.
\end{itemize}

For the experimental uncertainties, we consider the following sources:
\begin{itemize}
\item {Integrated luminosity:} The integrated luminosity measurements for the 2016, 2017, and 2018 data sets carry individual uncertainties between 0.82 and 1.2\%~\cite{CMS-LUM-17-003, CMS-PAS-LUM-20-001}. When combining the full 2016--2018 period, the overall uncertainty is 0.73\%.
\item {Pileup:} In simulation, the number of pileup interactions is predicted using a total inelastic \pp cross section of 69.2\unit{mb}~\cite{CMS:2018mlc}. To estimate the uncertainty, this cross section is varied by $\pm$4.6\%, yielding alternative templates via event reweighting. This uncertainty is treated as fully correlated across the different data-taking periods.
\item {Pileup jet identification:} A scale factor correction is applied to the simulation to match the efficiencies observed in data. It is used for the application of the PU jet ID, and its uncertainty is taken into account~\cite{puId} and is correlated across the years.
{\tolerance=1200
\item {Jet energy scale:} The jet energy scale uncertainties are split into 19 distinct sources~\cite{JET}. Additionally, separate variations are applied based on the generated jet type, namely \PQb, \PQc, light-flavor jets from \PW boson decay, and all other jets. The corresponding distributions are obtained by rescaling the jet momenta in the simulation. The correlations among the data-taking periods are evaluated for each source.
\par}
\item {Jet energy resolution:} Uncertainties in the jet energy resolution are treated separately for jets in the barrel and endcap regions by varying the resolution corrections to the simulation within their uncertainties. These uncertainties are treated as uncorrelated across the different data-taking periods.
\item {Missing transverse momentum:} The dominant source of uncertainty in \ptmiss stems from the jet energy calibration. Thus, \ptmiss is recalculated every time the jet momenta are rescaled. An additional contribution to this uncertainty, originating from particles that do not belong to the selected jets, is also estimated~\cite{CMS:2019ctu}. This uncertainty is treated as uncorrelated across the different data-taking periods.
\item {Lepton efficiencies:} Uncertainties in the reconstruction and trigger efficiencies for electrons and muons are determined as detailed in Ref.~\cite{TNP}. For each lepton flavor, both statistical and systematic uncertainties in the scale factors are included: the statistical uncertainties are uncorrelated across the different data-taking periods, whereas the systematic uncertainties are correlated. An overall normalization uncertainty of 0.5\% is additionally assigned to account for differences between DY and \ttbar events.
\item {\PQb tagging:} The \PQb tagging uncertainties are incorporated through variations of the scale factors~\cite{Bols_2020, CMS-DP-2023-005}. These variations are implemented by recalculating the event probability using all jets and their true flavor. The variations for the tight, medium, and loose selections are treated as fully correlated. For \PQb and \PQc jets, the uncertainties are split into components, including statistical, jet energy corrections, and pileup, among others; here, the statistical part is uncorrelated across the different data-taking periods, while the others are treated as correlated. For light-flavor quark and gluon jets, the uncertainties are divided into correlated and uncorrelated components.
\item {Trigger prefiring:} During data-taking, a gradual timing shift in the ECAL first-level trigger inputs for $\abs{\eta}>2.0$, together with a similar effect in the muon trigger, led to an additional trigger inefficiency~\cite{CMS:2020cmk}. Correction factors, which are functions of \pt, $\eta$, and time, are derived from data and applied to the simulation, with their statistical uncertainties properly taken into account.
\end{itemize}

The uncertainties listed above are applied to the continuum \ttbar production. For single top quark production and the two signal models, all experimental uncertainties are applied, while only a subset of the theoretical uncertainties is considered: higher-order ME corrections, PDF variations, parton shower effects, and the leptonic decay fraction of \PQb hadrons. The higher-order ME correction uncertainties are treated as uncorrelated across processes, while all other uncertainties are considered fully correlated. For the two signal models, parameters such as the \ttbar binding energy and \Mtop are fixed to their nominal values and are not treated as sources of systematic uncertainty. For the combined multijet/EW templates, a discussion of the associated uncertainties is provided in Section~\ref{sec:background}. An additional unconstrained normalization factor is included for the continuum \ttbar production, allowing its overall yield to be determined directly from the data.

In addition to the above-mentioned uncertainty sources, each bin is assigned a nuisance parameter that accounts for the statistical uncertainty in the central prediction~\cite{Conway:2011in}. The systematic variations can also be influenced by statistical fluctuations, particularly when these are derived from statistically independent simulations, the random fluctuations can match or even surpass the size of the systematic effect itself. To prevent these fluctuations from distorting the uncertainty estimates, a three-dimensional smoothing procedure~\cite{LOWESS, Cleveland01091988} is applied across the three axes of the histogram defined in Section~\ref{sec:signalfit} for each uncertainty source. This reduces the risk of unphysical constraints on the nuisance parameters while maintaining a robust treatment of uncertainties.

\section{Results}
\label{sec:results}

\begin{figure*}[!htp]
  \centering
  \includegraphics[width=\textwidth]{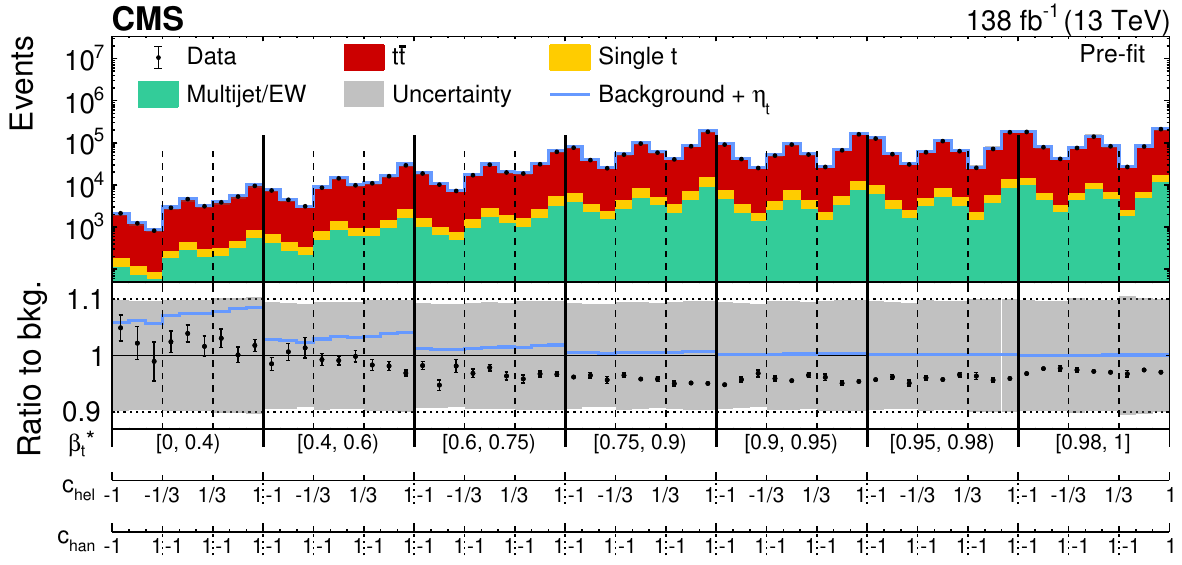} \\
  \includegraphics[width=\textwidth]{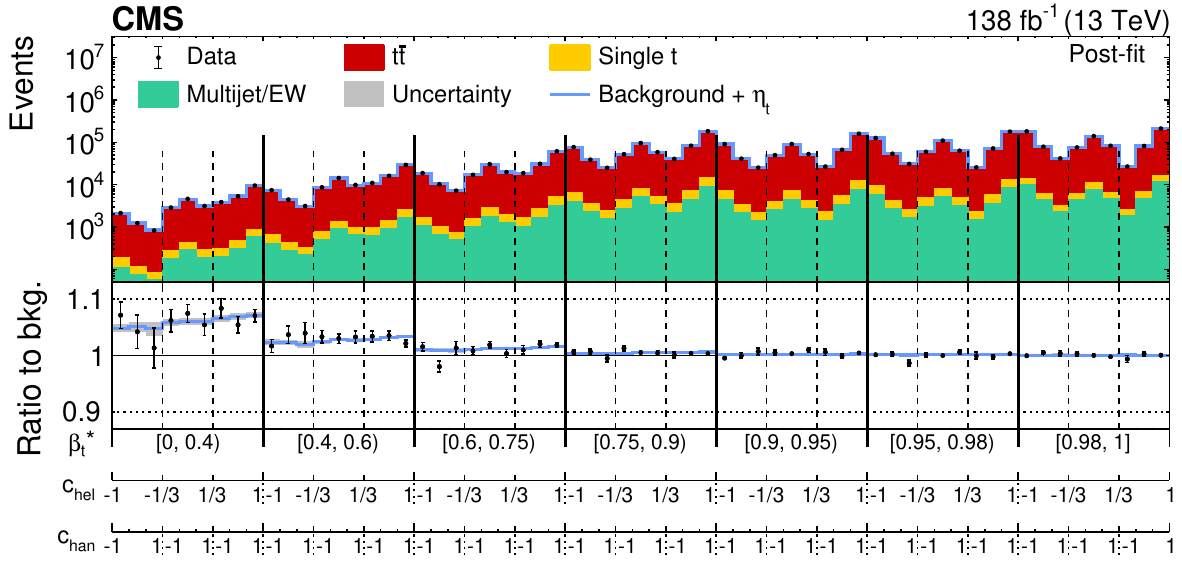}
  \caption{Pre-fit (upper) and post-fit (lower) $\boost\times\chel\times\chan$ distributions with all signal categories and data-taking periods combined. The data (points) are compared with the predictions without \etat (stacked histograms) and those including it (blue line). The $x$ axis shows bins of the unrolled three-dimensional distribution. The gray uncertainty band represents the combined statistical and systematic uncertainties in the predictions, while the vertical bars on the points indicate the statistical uncertainties in the data. The ratios to the predicted yields excluding \etat are shown in the lower panels.}
  \label{fig:signalfit_1}
\end{figure*}

\begin{figure*}[!p]
\centering
\includegraphics[width=0.48\textwidth]{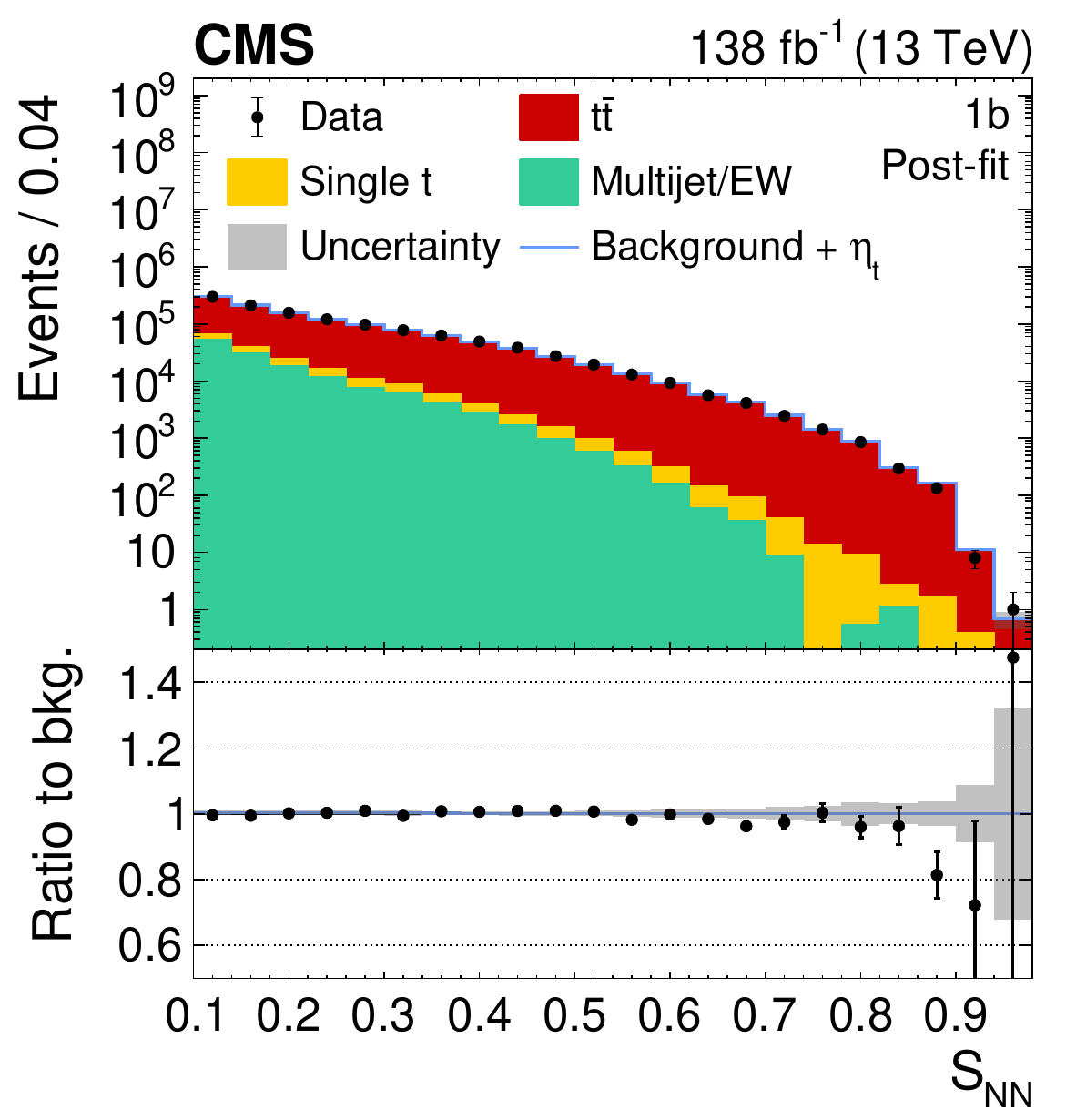}%
\hfill%
\includegraphics[width=0.48\textwidth]{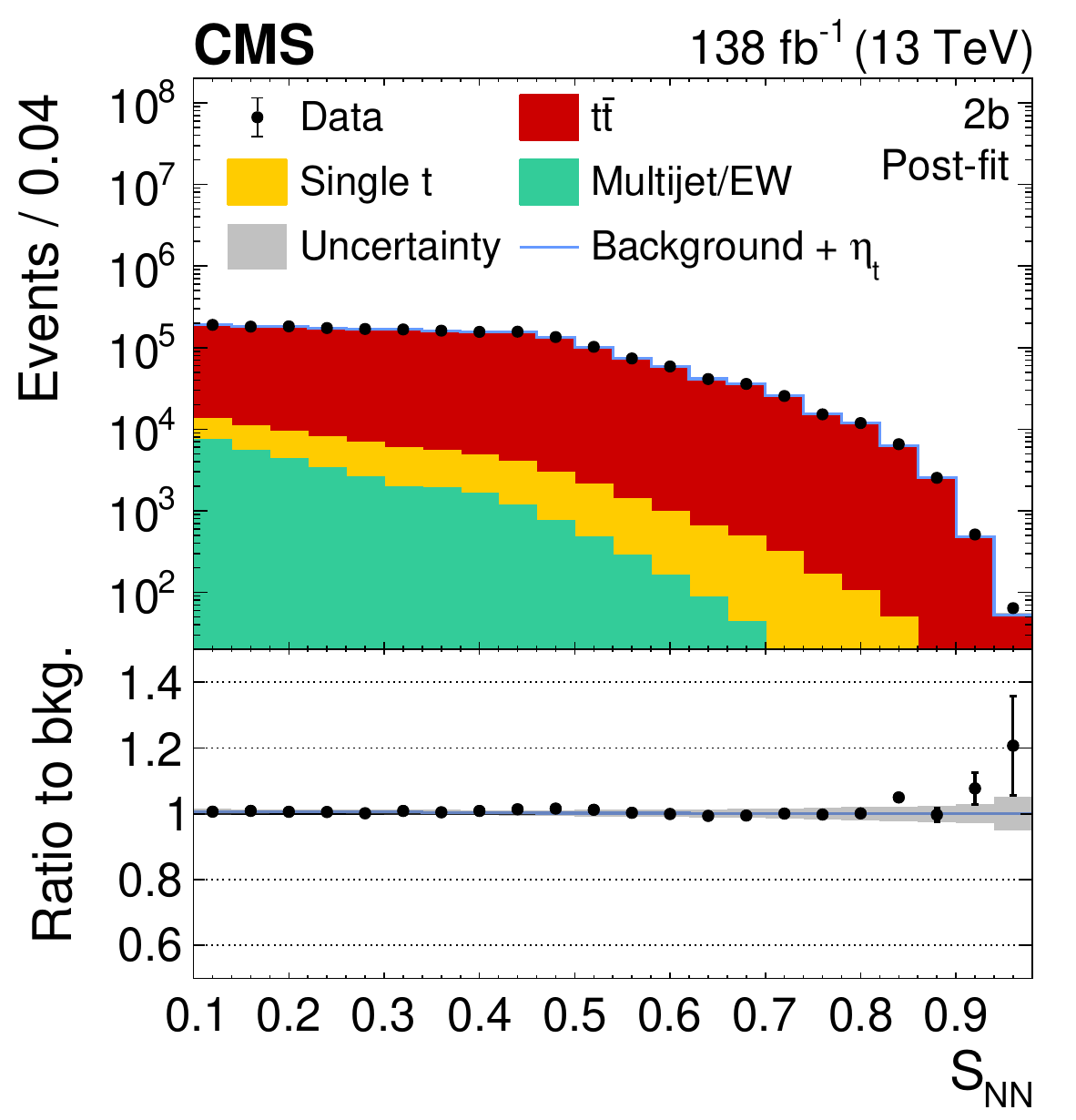}
\caption{Post-fit \SNN distributions for the 1\PQb (left) and 2\PQb (right) categories. The data (points) are compared with the predictions without \etat (stacked histograms) and those including it (blue line). The gray uncertainty band represents the post-fit combined statistical and systematic uncertainties in the predictions, while the vertical bars on the points indicate the statistical uncertainties in the data. The ratios to the predicted yields excluding \etat are shown in the lower panels.}
\label{fig:nn_postfit}
\end{figure*}

In Fig.~\ref{fig:signalfit_1}, the pre- and post-fit $\boost\times\chel\times\chan$ distributions are shown with all signal categories and data-taking periods combined. While the pre-fit distributions exhibit minor discrepancies between the data and the prediction, the post-fit distributions show excellent agreement, indicating that the unconstrained continuum \ttbar normalization factor and the systematic uncertainties effectively account for the observed differences. The reduction in the uncertainty band seen in the post-fit distribution, compared to the pre-fit, is largely driven by the correlations between nuisance parameters determined by the fit. Both pre- and post-fit distributions show an excess of events relative to the background at low \boost, where signal contributions are expected. A goodness-of-fit test based on the saturated model~\cite{BAKER1984437} results in a $p$-value~\cite{Cowan:2010js} of 0.59, indicating that the fitted model is consistent with the data. The post-fit distributions of \SNN are shown in Fig.~\ref{fig:nn_postfit}. The data and the predictions agree well, showing that the systematic uncertainties also account for the discrepancies observed in Fig.~\ref{fig:reconstruction_1}.

\begin{figure*}[!ph]
  \centering
  \includegraphics[width=0.8\textwidth]{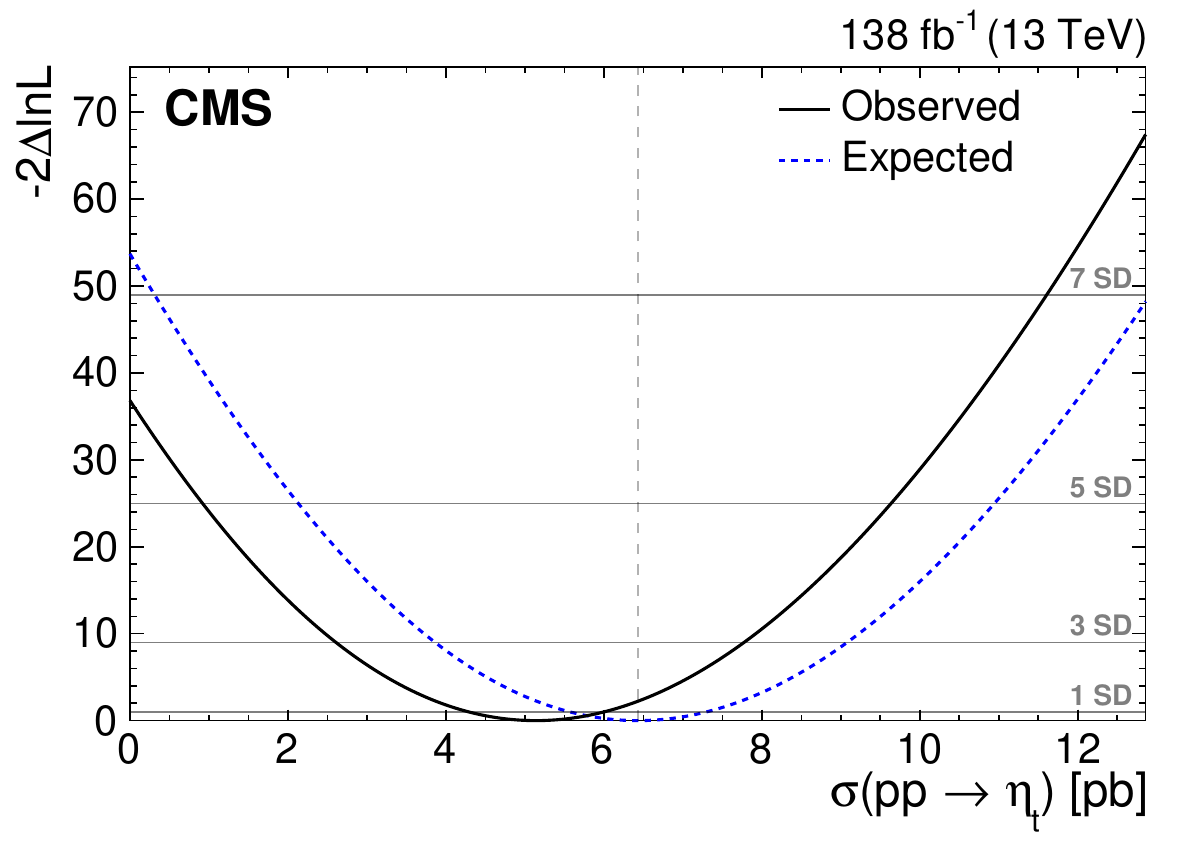}
  \caption{Observed (black solid line) and expected (blue dashed line) negative log-likelihood ratio scan as a function of the total \etat cross section. The gray horizontal lines represent the confidence intervals at the indicated SD levels. The vertical dashed line marks the nominal signal hypothesis ($\mu=1$), corresponding to a cross section of 6.43\unit{pb}.}
  \label{fig:LHscan}
\end{figure*}

\begin{figure*}[!ht]
  \centering
  \includegraphics[width=\textwidth]{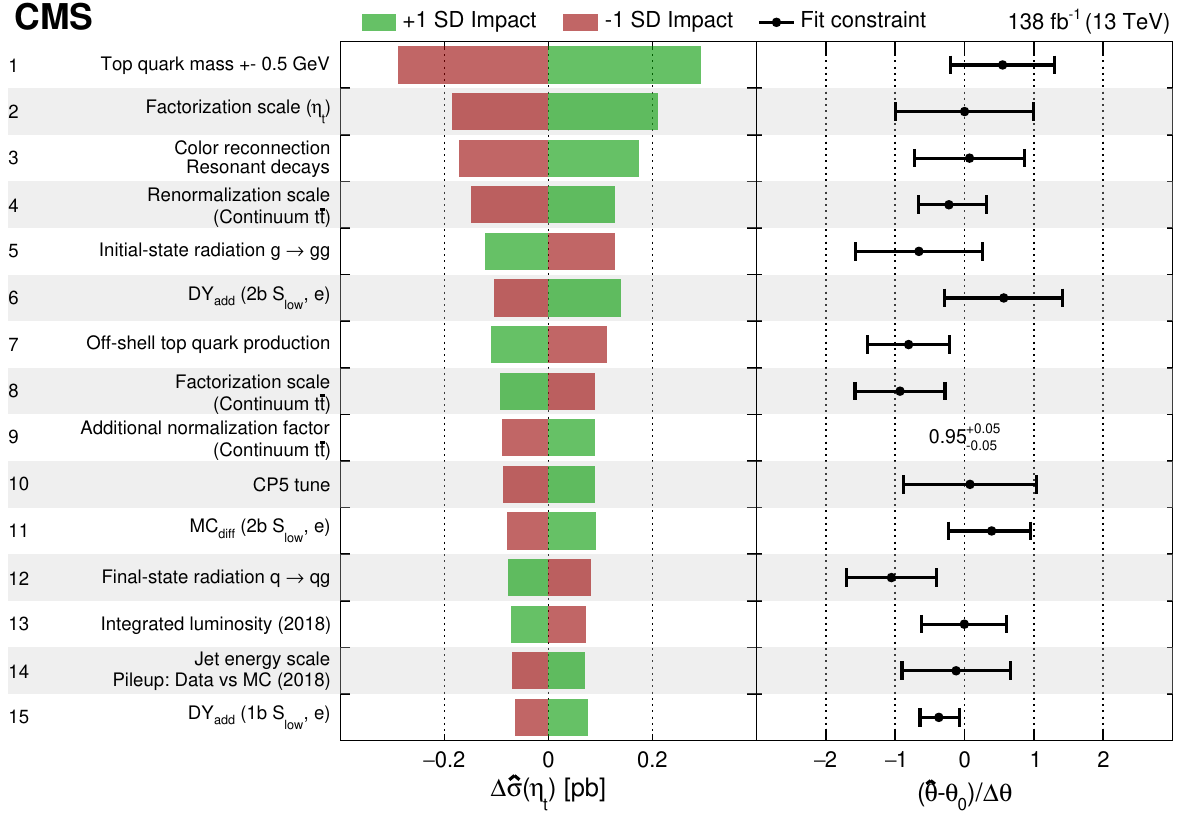}
  \caption{Impact of the 15 most significant systematic uncertainties on the measured \etat cross section $\widehat{\sigma}(\etat)$. The left panel shows the impact $\Delta\widehat{\sigma}(\etat)$ on the measurement when varying each nuisance parameter by $\pm1$ SD of its pre-fit uncertainty, represented by green and red bars, respectively. The right panel displays the pulls of the nuisance parameters, defined as $(\widehat{\theta}-\theta_0)/\Delta\theta$, where $\widehat{\theta}$ is the post-fit value, $\theta_0$ is the pre-fit value, and $\Delta\theta$ is the pre-fit uncertainty.}
  \label{fig:impacts}
\end{figure*}

To estimate the uncertainty in the measured cross section and the significance of the observation, we perform a negative log-likelihood ratio scan as a function of $\mu$. The scan result with $\mu$ converted into the corresponding cross section is presented in Fig.~\ref{fig:LHscan} for both the observed data and the Asimov expectation~\cite{Cowan:2010js}. The background-only hypothesis ($\mu=0$) is excluded with an observed (expected) significance of \etatsigobs (\etatsigexp) SDs. The best fit signal strength is measured to be $\mu=\mufitobs$, corresponding to a total \etat cross section of \etatxsec. The impact of the 15 most significant systematic uncertainties on the measured \etat cross section is shown in Fig.~\ref{fig:impacts}. The measurement is primarily sensitive to the modeling of continuum \ttbar and \etat. The leading sources of systematic uncertainty include the value of \Mtop, the \muf for \etat production, the \mur for \ttbar production, as well as the modeling of color reconnection and initial-state radiation. Overall, the nuisance parameters are well behaved; their post-fit values show no significant deviations from the pre-fit expectations, and only small to moderate constraints are observed for individual parameters, as shown in Fig.~\ref{fig:impacts}.

\begin{figure*}[!htp]
  \centering
  \includegraphics[width=0.495\textwidth]{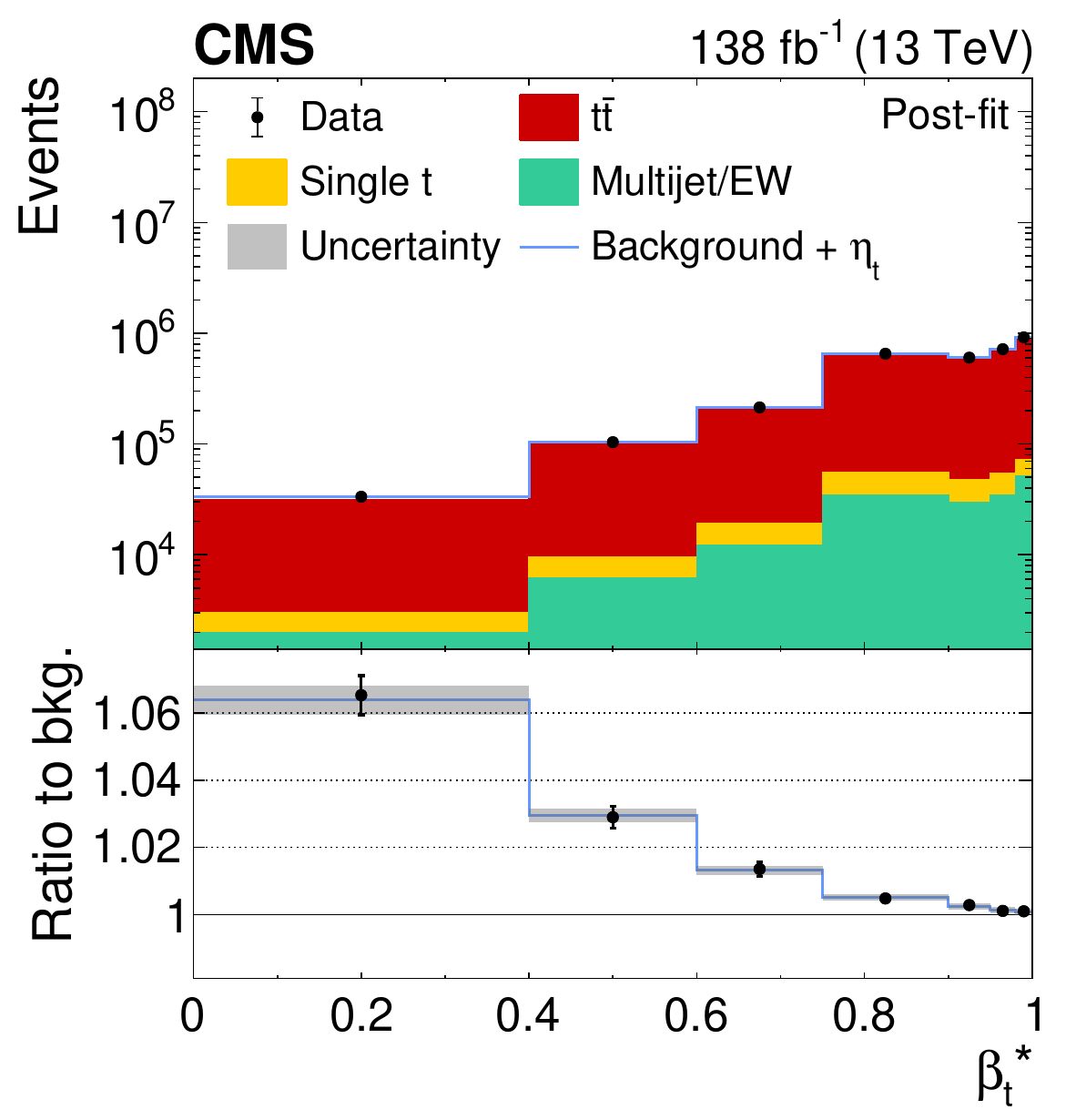}%
  \hfill%
  \includegraphics[width=0.495\textwidth]{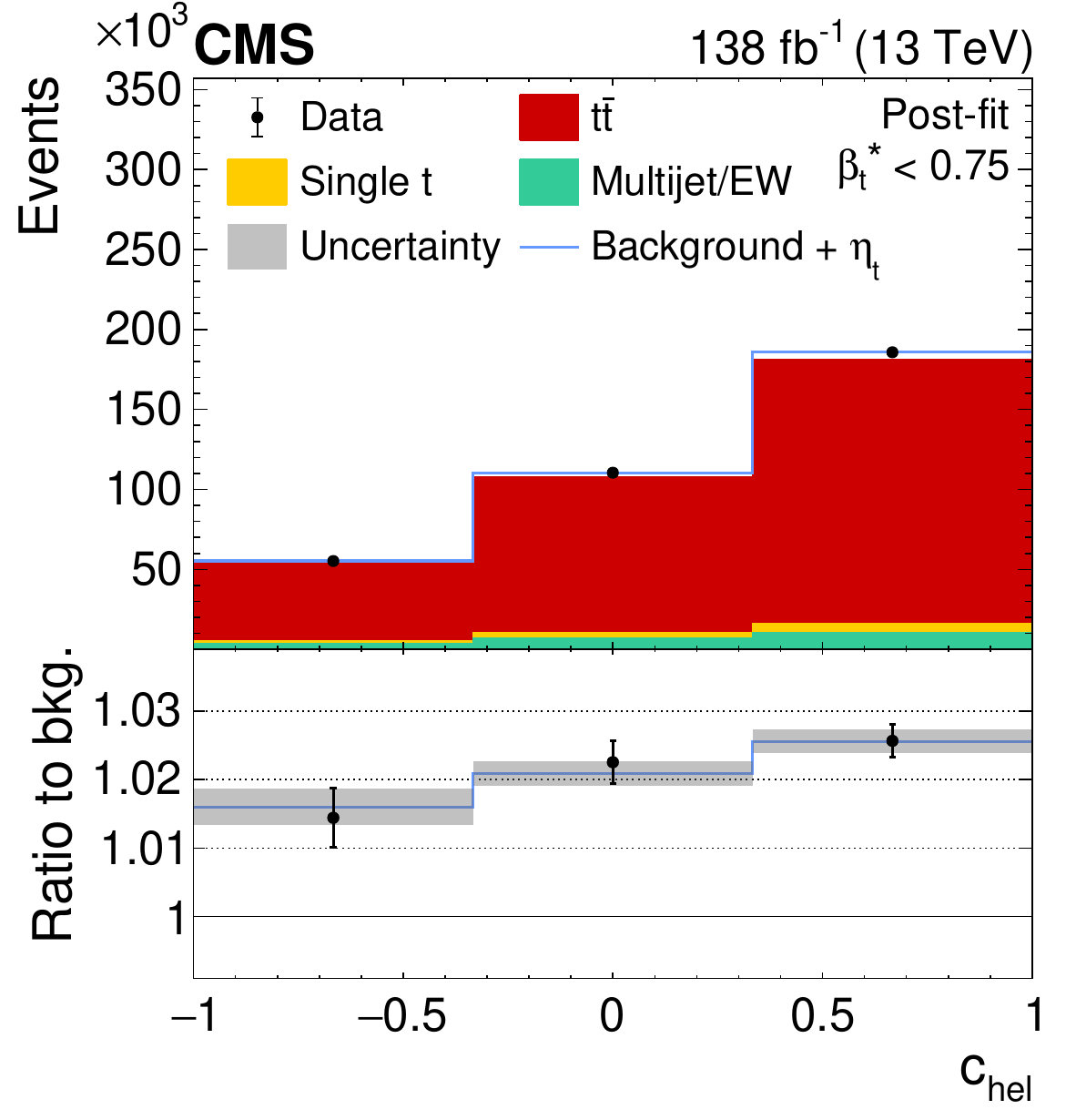} \\
  \includegraphics[width=0.495\textwidth]{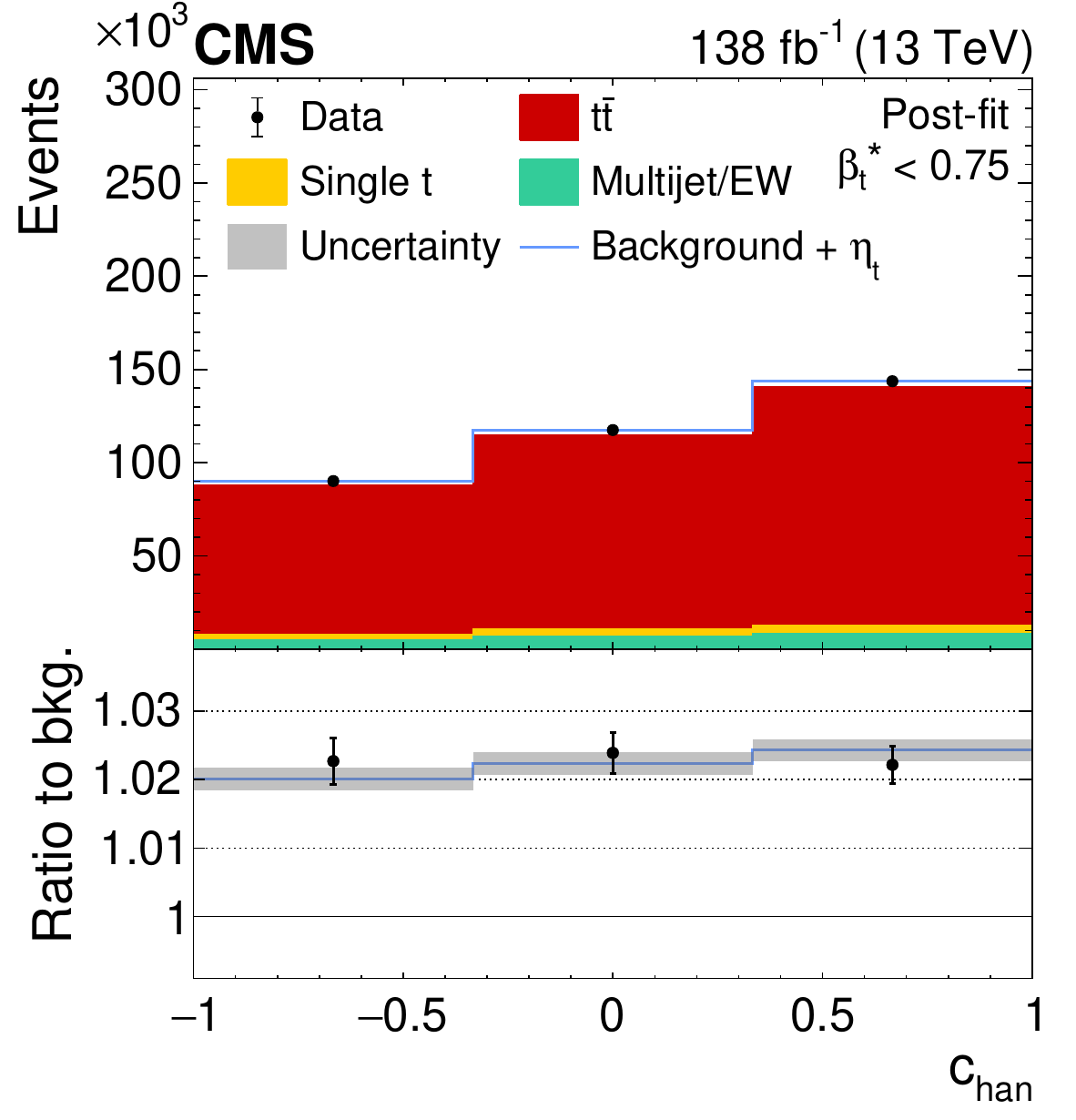}
  \caption{Post-fit distributions of \boost (upper left), \chel (upper right), and \chan (lower) for all four signal categories combined. The \boost distribution is integrated over \chel and \chan, while the \chel (\chan) distribution is integrated over \chan (\chel) and the first three bins of \boost ($\boost<0.75$). The data (points) are compared to the predictions without \etat (stacked histograms) and those including it (blue line). The gray uncertainty band represents the combined statistical and systematic uncertainties in the prediction. The vertical bars on the points indicate the statistical uncertainty in the data. The ratios to predicted yields excluding \etat are shown in the lower panels.}
  \label{fig:postfit_comb}
\end{figure*}

In Fig.~\ref{fig:postfit_comb} we present the post-fit distributions of \boost, \chel, and \chan, with the latter two restricted to the $\boost<0.75$ region. The excess at low \boost and the steeper slope in \chel observed in the data relative to the background-only prediction are consistent with the presence of the \etat signal, indicated by the blue line. In the \chan distribution, the predicted slope effect is less pronounced than in \chel; as a result, while the data are consistent with the \etat signal prediction, the slope is masked by statistical fluctuations.

For comparison, the signal extraction is also performed with \ttm instead of \boost, resulting in an expected significance of only 2.1 SDs. This demonstrates the substantial improvement from using \boost for the measurement. The expected significance using \boost alone, without \chel and \chan, is 6.3 SDs. The ambiguity in the identification of the down-type quark limits the sensitivity gain from including angular observables in the \lpj channel compared to the dilepton channel~\cite{TOP-24-007}.

An alternative fit using the \ttnrqcd signal model in place of \etat is performed. In this case, the background-only hypothesis is excluded with an observed (expected) significance of \ttnrqcdsigobs (\ttnrqcdsigexp) SDs. The best fit signal strength is determined to be $\mu=\ttnrqcdmufitobs$, corresponding to a total \ttnrqcd cross section of \ttnrqcdxsec. The discrepancy in the expected significances and the total measured cross sections between the two models is primarily due to the $\ttm<350\GeV$ cutoff in the \ttnrqcd model as discussed in Section~\ref{sec:samples}. Approximately 62\% of the parton-level \etat events fall below this threshold; scaling the measured \etat cross section by this fraction yields an estimated contribution of 3.2\unit{pb} in this mass range, indicating compatibility with the \ttnrqcd measurement.

\begin{figure*}[!htp]
  \centering
  \includegraphics[width=\textwidth]{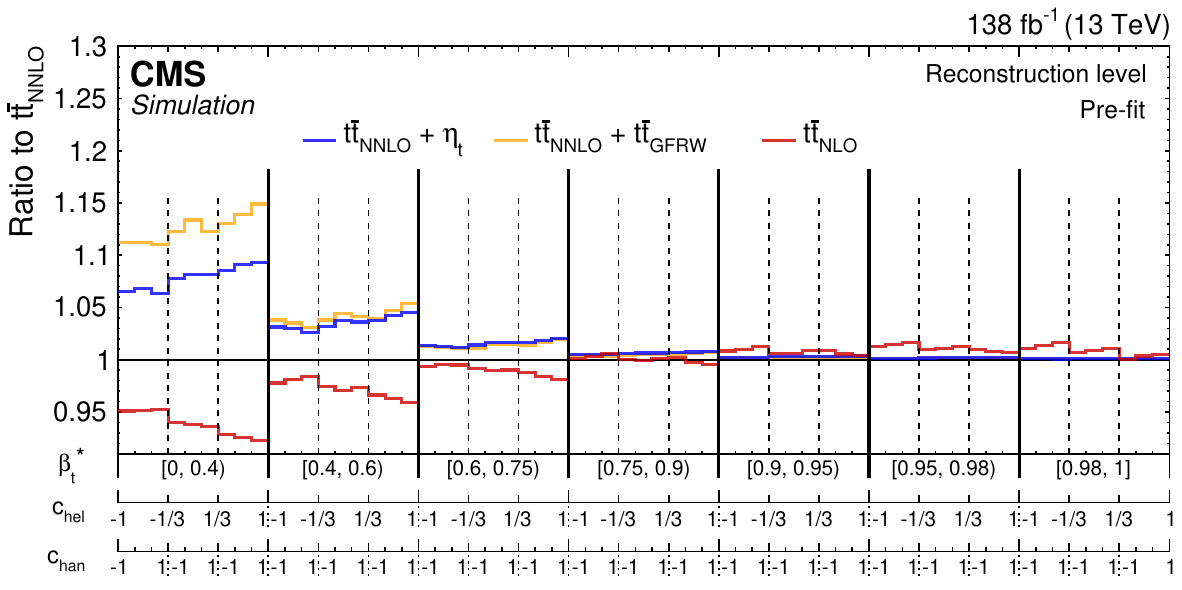}
  \caption{Pre-fit comparison of the \etat (blue) and \ttnrqcd (orange) models in the $\boost\times\chel\times\chan$ binning used for the signal extraction for all four signal categories combined, shown as ratios to the continuum \ttbar at NNLO QCD accuracy. The continuum \ttbar at NLO QCD accuracy normalized to the NNLO total cross section (red) is also shown for reference.}
  \label{fig:reco_model_compare}
\end{figure*}

A pre-fit comparison of the two models is presented in Fig.~\ref{fig:reco_model_compare}, showing the ratios relative to the continuum \ttbar at NNLO QCD accuracy in the $\boost\times\chel\times\chan$ binning. As illustrated in the figure, in the lowest \boost bins, the enhancement predicted by the \etat and \ttnrqcd models relative to the NNLO prediction is comparable in magnitude to the difference between the NLO and NNLO calculations; therefore, it is important to establish which baseline prediction is used to evaluate the excess. Due to the $\ttm<350\GeV$ cutoff in the \ttnrqcd model, its excess is more concentrated in the lowest \boost bin compared to the \etat model. The goodness-of-fit test with the \ttnrqcd model yields a $p$-value of 0.57. The similarity of this $p$-value to that of the \etat model, along with the comparable observed significances, suggests that both models are similarly compatible with the data.

\section{Summary}
\label{sec:summary}

A search for color singlet pseudoscalar toponium quasi-bound states has been conducted using proton-proton collision data collected by the CMS experiment from 2016 to 2018 at $\sqrt{s}=13\TeV$. While previous studies utilized the dilepton channel, this analysis performs the search in the \lpj channel for the first time. The strategy leverages two angular observables sensitive to the spin correlations and parity of the top quark-antiquark (\ttbar) system, together with \boost, defined as the magnitude of the velocity of one top quark in the rest frame of the other. Compared to the invariant mass of the \ttbar system \ttm, the \boost variable is more sensitive to the signal and more robust against specific experimental and modeling uncertainties. The background-only hypothesis is excluded with an observed (expected) significance of \etatsigobs (\etatsigexp) standard deviations. When using a simplified model simulating a pseudoscalar color singlet particle (\etat), the total signal production cross section is measured to be \etatxsec with the main uncertainties from the modeling of continuum \ttbar and \etat production. This result provides an independent confirmation of the excess reported in the dilepton channel. In the $\ttm<350\GeV$ mass range, results obtained with both the \etat model and an alternative model based on a Green's function reweighting of continuum \ttbar yield a consistent enhancement of roughly 3\unit{pb}. It is important to note that this study relies solely on approximate models, and further theoretical investigations are needed to more precisely describe the \ttbar threshold region.

\begin{acknowledgments}
We congratulate our colleagues in the CERN accelerator departments for the excellent performance of the LHC and thank the technical and administrative staffs at CERN and at other CMS institutes for their contributions to the success of the CMS effort. In addition, we gratefully acknowledge the computing centers and personnel of the Worldwide LHC Computing Grid and other centers for delivering so effectively the computing infrastructure essential to our analyses. Finally, we acknowledge the enduring support for the construction and operation of the LHC, the CMS detector, and the supporting computing infrastructure provided by the following funding agencies: SC (Armenia), BMFWF and FWF (Austria); FNRS and FWO (Belgium); CNPq, CAPES, FAPERJ, FAPERGS, and FAPESP (Brazil); MES and BNSF (Bulgaria); CERN; CAS, MoST, and NSFC (China); MINCIENCIAS (Colombia); MSES and CSF (Croatia); RIF (Cyprus); SENESCYT (Ecuador); ERC PRG and PSG, TARISTU24-TK10 and MoER TK202 (Estonia); Academy of Finland, MEC, and HIP (Finland); CEA and CNRS/IN2P3 (France); SRNSF (Georgia); BMFTR, DFG, and HGF (Germany); GSRI (Greece); MATE and NKFIH (Hungary); DAE and DST (India); IPM (Iran); SFI (Ireland); INFN (Italy); MSIT and NRF (Republic of Korea); MES (Latvia); LMTLT (Lithuania); MOE and UM (Malaysia); BUAP, CINVESTAV, CONACYT, LNS, SEP, and UASLP-FAI (Mexico); MOS (Montenegro); MBIE (New Zealand); PAEC (Pakistan); MSHE, NSC, and NAWA (Poland); FCT (Portugal); MESTD (Serbia); MICIU/AEI and PCTI (Spain); MOSTR (Sri Lanka); Swiss Funding Agencies (Switzerland); MST (Taipei); MHESI (Thailand); TUBITAK and TENMAK (T\"{u}rkiye); NASU (Ukraine); STFC (United Kingdom); DOE and NSF (USA).

\begin{sloppypar}
\setlength\emergencystretch{\hsize}
\hyphenation{Rachada-pisek} Individuals have received support from the Marie-Curie program and the European Research Council and Horizon 2020 Grant, contract Nos.\ 675440, 724704, 752730, 758316, 765710, 824093, 101115353, 101002207, 101001205, and COST Action CA16108 (European Union); the Leventis Foundation; the Alfred P.\ Sloan Foundation; the Alexander von Humboldt Foundation; the Science Committee, project no. 22rl-037 (Armenia); the Fonds pour la Formation \`a la Recherche dans l'Industrie et dans l'Agriculture (FRIA) and Fonds voor Wetenschappelijk Onderzoek contract No. 1228724N (Belgium); the Beijing Municipal Science \& Technology Commission, No. Z191100007219010, the Fundamental Research Funds for the Central Universities, the Ministry of Science and Technology of China under Grant No. 2023YFA1605804, the Natural Science Foundation of China under Grant No. 12535004, and USTC Research Funds of the Double First-Class Initiative No.\ YD2030002017 (China); the Ministry of Education, Youth and Sports (MEYS) of the Czech Republic; the Shota Rustaveli National Science Foundation (Georgia); the Deutsche Forschungsgemeinschaft (DFG), among others, under Germany's Excellence Strategy -- EXC 2121 ``Quantum Universe" -- 390833306, and under project number 400140256 - GRK2497; the Hellenic Foundation for Research and Innovation (HFRI), Project Number 2288 (Greece); the Hungarian Academy of Sciences, the New National Excellence Program - \'UNKP, the NKFIH research grants K 131991, K 138136, K 143460, K 143477, K 147557, K 146913, K 146914, K 147048, TKP2021-NKTA-64, and 2025-1.1.5-NEMZ\_KI-2025-00004, and MATE KKP and KKPCs Research Excellence and Flagship Research Groups grants (Hungary); the Council of Science and Industrial Research, India; ICSC -- National Research Center for High Performance Computing, Big Data and Quantum Computing, FAIR -- Future Artificial Intelligence Research, and CUP I53D23001070006 (Mission 4 Component 1), funded by the NextGenerationEU program, the Italian Ministry of University and Research (MUR) under Bando PRIN 2022 -- CUP I53C24002390006, PRIN PRIMULA 2022RBYK7T (Italy); the Latvian Council of Science; the Ministry of Science and Higher Education, project no. 2022/WK/14, and the National Science Center, contracts Opus 2021/41/B/ST2/01369, 2021/43/B/ST2/01552, 2023/49/B/ST2/03273, and the NAWA contract BPN/PPO/2021/1/00011 (Poland); the Funda\c{c}\~ao para a Ci\^encia e a Tecnologia (Portugal); the National Priorities Research Program by Qatar National Research Fund;  MICIU/AEI/10.13039/501100011033, ERDF/EU, ``European Union NextGenerationEU/PRTR", projects PID2022-142604OB-C21, PID2022-139519OB-C21, PID2023-147706NB-I00, PID2023-148896NB-I00, PID2023-146983NB-I00, PID2023-147115NB-I00, PID2023-148418NB-C41, PID2023-148418NB-C42, PID2023-148418NB-C43, PID2023-148418NB-C44, PID2024-158190NB-C22, RYC2021-033305-I, RYC2024-048719-I, CNS2023-144781, CNS2024-154769 and Plan de Ciencia, Tecnolog{\'i}a e Innovaci{\'o}n de Asturias, Spain; the Chulalongkorn Academic into Its 2nd Century Project Advancement Project, the National Science, Research and Innovation Fund program IND\_FF\_68\_369\_2300\_097, and the Program Management Unit for Human Resources \& Institutional Development, Research and Innovation, grant B39G680009 (Thailand); the Eric \& Wendy Schmidt Fund for Strategic Innovation through the CERN Next Generation Triggers project under grant agreement number SIF-2023-004; the Kavli Foundation; the Nvidia Corporation; the SuperMicro Corporation; the Welch Foundation, contract C-1845; and the Weston Havens Foundation (USA).
\end{sloppypar}
\end{acknowledgments}\section*{Data availability} Release and preservation of data used by the CMS Collaboration as the basis for publications is guided by the  \href{https://doi.org/10.7483/OPENDATA.CMS.1BNU.8V1W}{CMS data preservation, re-use and open access policy}.
\bibliography{auto_generated}
\cleardoublepage \appendix\section{The CMS Collaboration \label{app:collab}}\begin{sloppypar}\hyphenpenalty=5000\widowpenalty=500\clubpenalty=5000\cmsinstitute{Yerevan Physics Institute, Yerevan, Armenia}
{\tolerance=6000
A.~Gevorgyan\cmsorcid{0000-0003-2751-9489}, A.~Hayrapetyan, V.~Makarenko\cmsorcid{0000-0002-8406-8605}, A.~Tumasyan\cmsAuthorMark{1}\cmsorcid{0009-0000-0684-6742}
\par}
\cmsinstitute{Marietta Blau Institute for Particle Physics, Vienna, Austria}
{\tolerance=6000
W.~Adam\cmsorcid{0000-0001-9099-4341}, L.~Benato\cmsorcid{0000-0001-5135-7489}, T.~Bergauer\cmsorcid{0000-0002-5786-0293}, M.~Dragicevic\cmsorcid{0000-0003-1967-6783}, S.~Gundacker\cmsorcid{0000-0003-2087-3266}, A.K.~Guven\cmsorcid{0009-0004-5670-5138}, P.S.~Hussain\cmsorcid{0000-0002-4825-5278}, M.~Jeitler\cmsAuthorMark{2}\cmsorcid{0000-0002-5141-9560}, N.~Krammer\cmsorcid{0000-0002-0548-0985}, A.~Li\cmsorcid{0000-0002-4547-116X}, D.~Liko\cmsorcid{0000-0002-3380-473X}, M.~Matthewman, A.~Pfeiffer\cmsorcid{0000-0001-5328-448X}, J.~Schieck\cmsAuthorMark{2}\cmsorcid{0000-0002-1058-8093}, R.~Sch\"{o}fbeck\cmsAuthorMark{2}\cmsorcid{0000-0002-2332-8784}, M.~Shooshtari\cmsorcid{0009-0004-8882-4887}, M.~Sonawane\cmsorcid{0000-0003-0510-7010}, N.~Van~Den~Bossche\cmsorcid{0000-0003-2973-4991}, W.~Waltenberger\cmsorcid{0000-0002-6215-7228}, C.E.~Wulz\cmsAuthorMark{2}\cmsorcid{0000-0001-9226-5812}
% ...
\par}
$^{1}$Also at Yerevan State University, Yerevan, Armenia\\
$^{2}$Also at Technische Universit\"{a}t Wien, Vienna, Austria\\
% ...
\end{sloppypar}
\end{document}